\documentclass[12pt]{article}
\usepackage{multirow}
\usepackage{booktabs}
\usepackage{dsfont}
\usepackage{caption}
\usepackage{amssymb}
\usepackage{amsmath}
\usepackage{amstext}
\usepackage[authoryear,round]{natbib}
\usepackage{amsthm}
\usepackage{mathrsfs}
\usepackage{graphics} 
\usepackage{rotating}
\usepackage{afterpage}
\usepackage{color,verbatim}
\usepackage[bookmarks=false,
            colorlinks,
            linkcolor=blue,
            anchorcolor=blue,
            citecolor=blue,
            urlcolor=black
            ]{hyperref}
\usepackage{mathtools}
\usepackage{enumitem}
\usepackage{algorithm}
\usepackage{algpseudocode}
\usepackage{chngcntr}
\usepackage{xr}
\usepackage{subcaption}
\makeatletter
\newcommand*{\addFileDependency}[1]{%
  \typeout{(#1)}%
  \@addtofilelist{#1}
  \IfFileExists{#1}{}{\typeout{No file #1.}}
}

\selectfont

\def\squarebox#1{\hbox to #1{\hfill\vbox to #1{\vfill}}}
\def\boxit#1{\vbox{\hrule\hbox{\vrule\kern6pt
      \vbox{\kern6pt#1\kern6pt}\kern6pt\vrule}\hrule}}

\def\sumk0p{\sum_{k=0}^\ell}

\renewcommand{\hat}{\widehat}

\allowdisplaybreaks

\newtheoremstyle{break}
  {9pt}
  {12pt}
  {\upshape}
  {}
  {\bfseries}
  {.}
  {\newline}
  {}

\newtheorem{rem}{Remark}
\newtheorem{theo}{Theorem} 
\newtheorem{prop}{Proposition}

\theoremstyle{definition}
\newtheorem{exa}{Example}

\newtheorem{ass}{Assumption}
\newtheorem{cor}{Corollary}
\newtheorem{defi}{Definition}
\theoremstyle{remark}

\numberwithin{equation}{section}
\title{}
\author{}
\date{}

\begin{document}

\begin{center}
{\bf \Large 
Functional Causal Discovery via Conditional Covariance Ordering} \\[1cm]
\end{center}

\begin{center}
Keyu Li and Ruoxu Tan\footnote{Corresponding author: ruoxut@tongji.edu.cn} \\
School of Mathematical Sciences and School of Economics and Management, Tongji University 
\end{center} 

\linespread{1}\selectfont
\begin{abstract}
We study causal discovery where each node is a random function. Previous studies on this topic rely on structural assumptions, e.g., linearity or non-linearity, and distributional assumptions, e.g., Gaussianity or non-Gaussianity. In contrast, we make use of covariance operators to avoid these assumptions. Under functional additive noise models, we propose a new sufficient condition to identify a valid topological ordering based on comparing norms of conditional covariance operators. Taking advantage of this identifiability condition, we develop a new mixed regression model that subsumes linear and non-linear models. Together with variable selection, our procedure yields an estimation of the causal directed acyclic graph (DAG) for functional variables. In theory, we develop the least-squares-type theory of this regression model, and derive asymptotic consistency of order determination, sparse regression, as well as identifying the DAG. Computational algorithms based on discrete observations are provided. Applied to simulated data, our approach performs satisfactorily among existing approaches. A real data example of brain effective connectivity is also presented.
\end{abstract}


\linespread{1.7}\selectfont
\section{Introduction}\label{sc_in}
Multivariate functional data arise in clinical studies, neuroscience, and other applications in which multiple related processes are recorded as trajectories. Understanding the causal relationships among these processes is important for investigating mechanisms such as neural interactions and gene regulation~\citep{Reid2019Advancing,Ota2026}. A central challenge is to identify causal directions among multivariate functional variables without assuming a specific functional form or error distribution. 

There exists a rich literature on causal discovery for random variables, which can be roughly categorized into constraint-based approaches using conditional independence tests \citep{Kalisch2007,maathuis2009,Huang2020}, score-based approaches by minimizing a predefined score function \citep{Loh2014,Han2016,Zheng2018}, and asymmetry-driven approaches leveraging non-linear or non-Gaussian residuals \citep{Shimizu2006,Hoyer2009,Shimizu2011,Mitrovic2018}.  Each branch of work faces certain challenges. For example, approaches based on conditional independence tests typically only yield completed partially directed acyclic graphs, while asymmetry-driven approaches rely on specific structural or distributional assumptions.

It is non-trivial to extend the causal discovery techniques from random variables to random functions. Existing work has developed several approaches under different modeling assumptions. \citet{Lee2022} postulated a non-linear Gaussian model for functional data and established identification procedures based on a novel non-linear regression framework. Based on a functional linear non-Gaussian model, \citet{Yang2024} adopted functional principal component analysis to extend the LiNGAM theory to Hilbert spaces, while \citet{Zhou2023} developed a Bayesian approach to identify the DAG, which was later extended to allow cycles \citep{Roy2023}. Also, \citet{Lee2024} extended the classical PC algorithm to functional data. On the other hand, there exists a branch of studies focusing on undirected graphs for functional data \citep{Lee2016a,Li2018,Qiao2019,Lee2023}, which is fundamentally different from directed graphs; see \citet{Zhao2026} for a review. In the field of treatment effect estimation, there is also a growing interest involving functional variables \citep{Ecker2024,Tan2025,Wang2026}.
 
In this article, we investigate the causal identifiability for functional data beyond specific structural or distributional assumptions. Our first contribution is a new sufficient condition for recovering a valid topological ordering in functional additive noise models. Motivated by conditional-variance ordering for scalar variables \citep{Park2020}, we compare the Hilbert–Schmidt norms of integrated conditional covariance operators for functional variables. The new condition accommodates both linear and non-linear models with Gaussian or non-Gaussian errors. Notably, it is strictly more general than the assumption of bounded linear models with equal norm of error covariance operators and the edge non-degeneracy.

Taking advantage of this identifiability condition, our second contribution is a new regression framework that explicitly consists of linear and non-linear components. We follow the established strategy of estimating an ordering and then identifying edges through sparse regression \citep{Shojaie2010,Buhlmann2014,Lee2022,Wang2025}. The least-squares-type theory of this regression model has been established. We adapt a group SCAD penalty \citep{Fan2001} for variable selection. This regression framework can be of independent interest in the field of functional regression. 
 
In asymptotic analysis, we derive the uniform consistency of order determination, the convergence rate of the regression coefficients, and the consistency of variable selection, which together imply the consistency of the procedure of identifying the DAG under fully observed functions. We develop an estimation procedure for discretely observed data, which shows prominent performance compared with existing methods in simulation studies. The application of our approach to an electroencephalography dataset provides brain connectivity that is consistent with the related literature. 

The rest of the article is organized as follows.  Section~\ref{sc_fanm} introduces functional additive noise models and the identifiability result via the conditional covariance operator, while Section~\ref{sc_FMANM} develops the mixed regression model and its estimation. Sections~\ref{sc_theo} and~\ref{sc_comp} present asymptotic theory and computation from discrete observations, respectively. Section~\ref{sc_num} presents simulations and a brain effective connectivity application, followed by discussion in Section~\ref{sc_dis}. Proofs and additional experimental details are provided in the Supplementary Material.
 
\section{Functional Additive Noise Model and Its Identifiability}\label{sc_fanm} 

\subsection{Model and Notations} \label{subsc_fanm}
Let $\mathbb G=(\mathbb V,\mathbb E)$ be a directed acyclic graph (DAG), where $\mathbb V=\{1,\dots,p\}$ and $(i,j)\in\mathbb E$ represents the directed edge $i\to j$, under which $i$ is a parent of $j$ and $j$ is a child of $i$. For each node $i$, let $\mathrm{pa}(i;\mathbb G)$, $\mathrm{an}(i;\mathbb G)$, $\mathrm{de}(i;\mathbb G)$, and $\mathrm{nd}(i;\mathbb G)=\mathbb V\setminus(\{i\}\cup\mathrm{de}(i;\mathbb G))$ denote its parents, ancestors, descendants, and non-descendants, respectively. A topological ordering $\pi_{1\to p}=(\pi_1,\dots,\pi_p)$ is a permutation of $\mathbb V$ in which every parent precedes its children. We denote the set of all such orderings by $\Pi(\mathbb G)$.

For each $i\in\mathbb{V}$, $X_i$ is a real-valued random function on a compact interval, rescaled to $[0,1]$, and belongs to $\mathcal H_X=L^2([0,1])$, the Hilbert space with inner product $\langle f,g\rangle_{\mathcal H_X}=\int_0^1 f(t)g(t)\,dt$ and induced norm $\|\cdot\|_{\mathcal H_X}$. Let $\mathcal B(\mathcal H_X,\mathcal H_X)$ and $\mathcal B_{\mathrm{HS}}(\mathcal H_X,\mathcal H_X)$ denote the spaces of bounded linear operators and Hilbert--Schmidt operators from $\mathcal H_X$ to itself, respectively. For $A\in\mathcal B(\mathcal H_X,\mathcal H_X)$, its adjoint $A^*$ is the unique operator satisfying $\langle Af,g\rangle_{\mathcal H_X}=\langle f,A^*g\rangle_{\mathcal H_X}$, for $f,g\in\mathcal H_X$. A positive semidefinite operator $A\in\mathcal B(\mathcal H_X,\mathcal H_X)$ is trace-class if $ \mathrm{tr}(A) = \sum_{k=1}^{\infty} \langle Ae_k,e_k\rangle_{\mathcal H_X}<\infty$, where $\{e_k\}_{k=1}^{\infty}$ is any orthonormal basis of $\mathcal H_X$. An operator $A\in\mathcal B(\mathcal H_X,\mathcal H_X)$ is Hilbert--Schmidt if $\|A\|_{\mathrm{HS}}^2 := \mathrm{tr}(A^*A) = \sum_{k=1}^{\infty}\|Ae_k\|_{\mathcal H_X}^2 < \infty
$. It follows that $\mathcal B_{\mathrm{HS}}(\mathcal H_X,\mathcal H_X) \subseteq \mathcal B(\mathcal H_X,\mathcal H_X)$. Moreover, $\mathcal B_{\mathrm{HS}}(\mathcal H_X,\mathcal H_X)$ is a Hilbert space with inner product $ \langle A,B\rangle_{\mathrm{HS}} = \mathrm{tr}(A^*B) = \sum_{k=1}^{\infty} \langle Ae_k,Be_k\rangle_{\mathcal H_X}$, whereas $\mathcal B(\mathcal H_X,\mathcal H_X)$ is a Banach space with the operator norm $ \|A\|_{\mathrm{op}} = \sup_{\|f\|_{\mathcal H_X}=1}\|Af\|_{\mathcal H_X}$.  For $A\in\mathcal B(\mathcal H_X,\mathcal H_X)$, define its kernel and range by $ \ker(A) = \{f\in\mathcal H_X:Af=0\} $ and $ \mathrm{ran}(A) = \{Af:f\in\mathcal H_X\}$, respectively.  

\begin{defi}\label{defFANM}(Functional Additive Noise Models)
A random vector $\mathbf{X} = (X_1, \dots, X_p)^\top$ follows a functional additive noise model\,(FANM) with respect to a DAG $\mathbb{G}=(\mathbb{V},\mathbb{E})$ if, for each node $i \in \mathbb{V}$, there exist measurable operators $h_{j,i}$: $\mathcal{H}_{X} \to \mathcal{H}_{X} $ such that
\begin{align*}
X_i = \sum_{j \in \mathrm{pa}(i;\mathbb{G})} h_{j,i}(X_j) + \epsilon_i,
\end{align*}
where $\epsilon_1,...,\epsilon_p$ are mutually independent mean-zero random functions in $\mathcal{H}_X$. Here, an ordered pair $(j,i)$ belongs to $\mathbb E$ if and only if $h_{j,i}(X_j)$ is almost surely not a constant.
\end{defi}

We assume that $\mathbf{X}$ follows an FANM throughout. Our methodology largely relies on the notion of covariance operators. For $i, j \in \mathbb{V}$, let $\mu_i = E(X_i)$ denote the mean function of $X_i$, and let $\Gamma_{X_i X_j} = E\{(X_i - \mu_i) \otimes (X_j - \mu_j) \}$ denote the covariance operator,  where the tensor product operator $(X_i \otimes X_j): \mathcal{H}_{X} \to \mathcal{H}_{X}$ acts as $(X_i \otimes X_j)(f) = \langle X_i, f \rangle_{\mathcal{H}_{X}} X_j$ for $f \in \mathcal{H}_{X}$. In addition to the unconditional operators, we also define the (integrated and thus deterministic) conditional covariance operator $\Gamma_{X_i X_i \mid \mathbf{X}_\mathbb{S}}: \mathcal{H}_{X} \to \mathcal{H}_{X}$ by $\Gamma_{X_i X_i \mid \mathbf{X}_\mathbb{S}} = E[\{ X_i - E(X_i \mid \mathbf{X}_\mathbb{S}) \} \otimes \{ X_i - E(X_i \mid \mathbf{X}_\mathbb{S}) \} ]$, for $i \in \mathbb{V}$ and  $\mathbb{S} \subseteq \mathbb{V} \setminus \{i\}$. We assume $E\|X_i\|_{\mathcal{H}_X}^4<\infty$ for every $i\in\mathbb V$ throughout, and thus these operators are well defined. In particular, the (conditional) covariance operators are trace-class and hence Hilbert--Schmidt \citep{Hsing2015}. 

\subsection{Identifiability via Conditional Covariance Operators}\label{subsc_iden}
The identifiability of the causal order under our framework rests on the conditional covariance operator, which serves as a crucial basis for our methodology and theory. Since our methodology does not rely on conditional independence tests, we do not require the faithfulness assumption. Instead, we postulate the assumption of causal sufficiency, i.e., every common cause of any two or more variables in $\mathbb{V}$ is in $\mathbb{V}$, a standard assumption in the literature~\citep{maathuis2009,Shojaie2010,Han2016}.  

We call an ordering $\pi^{\min}_{1\to p} = (\pi^{\min}_1, \dots, \pi^{\min}_p)$ a \emph{minimized conditional covariance ordering} of $\mathbb{G}$ if it is obtained through the following sequential minimization:
\begin{equation}
    \pi^{\min}_i \in \arg\min_{j \in \mathbb{V} \setminus \mathbb{S}_{i}} \|\Gamma_{X_jX_j \mid \mathbf{X}_{\mathbb{S}_{i}}}\|_{\mathrm{HS}}, \quad \text{for } i = 1, \dots, p, \label{eq:2.1}
\end{equation}
where $\mathbb{S}_{i} = \{\pi^{\min}_1, \dots, \pi^{\min}_{i-1}\}$ for $i=2,\dots,p$ and $\mathbb{S}_1 = \emptyset$. If the minimizer in \eqref{eq:2.1} is not unique at some step, $\pi^{\min}_i$ is selected from the argmin set by any deterministic tie-breaking rule. 

For scalar variables, \citet{Park2020} showed that the minimized conditional variance ordering is a topological ordering if the forward or backward stepwise criterion is satisfied. Here, we provide a different sufficient condition to guarantee $\pi^{\min}_{1\to p}\in\Pi(\mathbb{G})$, which is formulated in terms of \emph{ancestral subsets}\,\citep{Richardson2003} and the inequality~\eqref{eq:2.2} below.  
A subset $\mathbb{S}\subseteq\mathbb{V}$ is called ancestral if for every node in $\mathbb{S}$, all its ancestors also belong to $\mathbb{S}$. 
\begin{theo}
\label{theorem1}
Suppose that for any ancestral subset $\mathbb S \subset \mathbb{V} $, any node $ j \in \mathbb{V} \setminus \mathbb S $ with $ \mathrm{pa}(j;\mathbb{G}) \subseteq \mathbb S $, and any descendant $ k \in \mathrm{de}(j;\mathbb{G})$, we have
\begin{equation}
    \|\Gamma_{X_j X_j \mid \mathbf{X}_\mathbb S}\|_{\mathrm{HS}} < \|\Gamma_{X_k X_k \mid \mathbf{X}_\mathbb S}\|_{\mathrm{HS}}.\label{eq:2.2}
\end{equation}
Then, any minimized conditional covariance ordering $\pi^{\min}_{1\to p}$ obtained from \eqref{eq:2.1} is a true topological ordering of $\mathbb{G}$, i.e., $\pi^{\min}_{1\to p} \in \Pi(\mathbb{G})$.
\end{theo}
\begin{rem}
Theorem~\ref{theorem1} remains valid with any other norm in place of the HS norm, provided the same norm is used in both~\eqref{eq:2.1} and~\eqref{eq:2.2}. The resulting ordering should be interpreted in terms of the chosen norm. We focus on the HS norm for its computational convenience, theoretical properties, and its common usage in the existing literature of functional graphical models  \citep{Lee2022,Lee2023,Lee2024}.
\end{rem}  

The proof of Theorem~\ref{theorem1} is provided in Section~A of the Supplementary Material. The idea of~\eqref{eq:2.2} is to compare how much variation remains unexplained after accounting for the variables already selected. When all parents of a node have been selected, only its own noise remains. Its descendants may still have unexplained variation from both their own noise and parental effects. Condition~\eqref{eq:2.2} requires this combined variation to give the descendants larger conditional covariance norms. Selecting the smallest norm therefore helps place causes before their effects. In practice, it is straightforward to find the minimized conditional covariance ordering, where we apply a greedy search that successively selects the node with the smallest conditional HS norm and any tie-breaking rule. We defer the detailed estimation to Section~\ref{subsc_est}.
\begin{prop}\label{prop_01}
Suppose that, in Definition~\ref{defFANM}, the operators $h_{j,i}$ are bounded and linear, and the error covariance operators have equal HS norm, i.e., $\|\Gamma_{\epsilon_i\epsilon_i}\|_{\mathrm{HS}}=c$, for all $i\in\mathbb V$ and some $c>0$. Suppose further that every edge is non-degenerate in the sense that $h_{i,j}\Gamma_{\epsilon_i\epsilon_i}h_{i,j}^*\neq 0$, for $(i,j)\in\mathbb E$. Then, for any ancestral subset $\mathbb S \subset \mathbb{V}$, any node $j \in \mathbb{V} \setminus \mathbb S$ with $\mathrm{pa}(j;\mathbb{G}) \subseteq \mathbb S $, and any descendant $ k \in \mathrm{de}(j;\mathbb{G})$, we have $\|\Gamma_{X_j X_j \mid \mathbf{X}_\mathbb S}\|_{\mathrm{HS}} < \|\Gamma_{X_k X_k \mid \mathbf{X}_\mathbb S}\|_{\mathrm{HS}}$.
\end{prop} 
The proof of Proposition~\ref{prop_01} is provided in Section~D of the Supplementary Material. In the literature of causal discovery for random variables, the identifiability has been established under linear models with equal error variance \citep{Peters2014,Chen2019}. The corresponding assumption for functional data is linear and bounded operators with equal norm of error covariance operators; while Proposition~\ref{prop_01} shows that this assumption with non-degenerate edges implies our condition at~\eqref{eq:2.2}. Moreover, we show that the converse is not true by providing a simple example where our condition at~\eqref{eq:2.2} is also able to identify models with unequal norms of error covariances.

\begin{exa}
Consider a DAG $\mathbb{G}: X_1 \rightarrow X_2 \rightarrow X_3$ and an orthonormal set $\{\phi_1,\phi_2\}$ in $\mathcal{H}_X$ (e.g., $\phi_1(t)=\sqrt{2}\sin(\pi t),\ \phi_2(t)=\sqrt{2}\sin(2\pi t)$ on $t\in[0,1]$). We define a FANM by $X_1(t) = \epsilon_1(t)$, $X_2(t) = \int_0^1 \beta_{2,1}(t,s)\,X_1(s)\,ds + \epsilon_2(t)$ and $X_3(t) = \int_0^1 \beta_{3,2}(t,s)\,X_2(s)\,ds + \epsilon_3(t)$, where $\beta_{2,1}(t,s) = 1.5\,\phi_1(t)\phi_1(s) + 0.8\,\phi_2(t)\phi_2(s)$ and $\beta_{3,2}(t,s) = 0.8\,\phi_1(t)\phi_1(s) + 1.5\,\phi_2(t)\phi_2(s)$, for $t,s\in[0,1]$. The error processes satisfy $\epsilon_i(t)=\epsilon_{i1}\phi_1(t)+\epsilon_{i2}\phi_2(t)$, where the coefficients $(\epsilon_{i1},\epsilon_{i2})$ are mutually independent and independent across $i$ with mean zero and $\operatorname{Var}(\epsilon_{i1})=\operatorname{Var}(\epsilon_{i2})=\sigma_i^2>0$. Here, $\sigma_i$ can be unequal, i.e., unequal norms of error covariances.

Direct computation gives $\|\Gamma_{X_1X_1}\|_{\mathrm{HS}} = \sqrt{2}\,\sigma_1^2$, $\|\Gamma_{X_2X_2}\|_{\mathrm{HS}} = \sqrt{(2.25\sigma_1^2 + \sigma_2^2)^2 + (0.64\sigma_1^2 + \sigma_2^2)^2}$, and $\|\Gamma_{X_3X_3}\|_{\mathrm{HS}} = \sqrt{(1.44\sigma_1^2 + 0.64\sigma_2^2 + \sigma_3^2)^2 + (1.44\sigma_1^2 + 2.25\sigma_2^2 + \sigma_3^2)^2}$. For any $\sigma_i^2>0$, we have $\|\Gamma_{X_2X_2}\|_{\mathrm{HS}}>\sqrt{2}\,\sigma_1^2$ and $\|\Gamma_{X_3X_3}\|_{\mathrm{HS}}>\sqrt{2}\,\sigma_1^2$. Hence the minimum of \eqref{eq:2.1} is attained at node $1$, giving $\pi_1^{\min}=1$, $\mathbb{S}_2=\{1\}$. Moreover, we have   $\|\Gamma_{X_2X_2\mid X_1}\|_{\mathrm{HS}} = \sqrt{(\sigma_2^2)^2 + (\sigma_2^2)^2} = \sqrt{2}\,\sigma_2^2$ and $\|\Gamma_{X_3X_3\mid X_1}\|_{\mathrm{HS}} = \sqrt{(0.64\sigma_2^2 + \sigma_3^2)^2 + (2.25\sigma_2^2 + \sigma_3^2)^2}$. For any $\sigma_i^2>0$, we have $\|\Gamma_{X_3X_3\mid X_1}\|_{\mathrm{HS}} > \sqrt{2}\,\sigma_2^2 = \|\Gamma_{X_2X_2\mid X_1}\|_{\mathrm{HS}}$, and thus we identify  $\pi_2^{\min}=2$, $\mathbb{S}_3=\{1,2\}$. Finally, only node $3$ remains, and thus we obtain the correct ordering $\pi^{\mathrm{min}}_{1\to 3}=(1,2,3)$.
\end{exa}
In the literature of functional DAG, the identifiability has been established under structural (e.g., linear or non-linear) or distributional (e.g., Gaussian or non-Gaussian) assumptions \citep{Lee2022,Zhou2023,Yang2024}. In contrast, our identifiability condition requires an ordering for norms of conditional covariances but does not postulate specific requirements on structure, distribution or error norms.   

After order determination, edge selection can be carried out through variable selection in sparse regression \citep{Shojaie2010,Buhlmann2014,Lee2022,Wang2025}. Following this strategy, we introduce our mixed regression models in Section~\ref{sc_FMANM}.

\section{Functional Mixed Additive Noise Model}\label{sc_FMANM}

\subsection{Model and Theoretical Foundation}\label{subsc_FMANM} 

To model non-linear relationships among multivariate random functions, \citet{Lee2022} introduced a nested RKHS framework constructed via a smooth nested kernel function $\kappa: \mathcal{H}_X \times \mathcal{H}_X \to \mathbb{R}$, defined as $\kappa(f, g) = \rho( \langle f, f \rangle_{\mathcal{H}_X}, \langle f, g \rangle_{\mathcal{H}_X}, \langle g, g \rangle_{\mathcal{H}_X})$, where $\rho: \mathbb{R}^3 \to \mathbb{R}$ is a suitably chosen function that ensures $\kappa$ is positive definite. The nested RKHS is defined as $\mathscr{H}_X = \overline{\mathrm{span}}\{\kappa(\cdot, f) : f \in \mathcal{H}_X\}$, which represents non-linear features of the original random functions. The non-linear component of our functional mixed additive noise model (FMANM) introduced below relies on this nested RKHS framework. In addition,  FMANM explicitly includes a linear component.

\begin{defi}[Functional Mixed Additive Noise Model]
A random vector $\mathbf{X}=(X_1,\dots,X_p)$ follows a functional mixed additive noise model with respect to a DAG $\mathbb{G}$ if for each node $i\in\mathbb{V}$,
\begin{align}\label{eq_FMANM}
X_i = \mu_i + \sum_{j\in\mathrm{pa}(i;\mathbb{G})} \bigl[ A_{i,j}^* (X_j - \mu_j) + B_{i,j}^* \{\kappa(\cdot, X_j) - m_j\} \bigr] + \epsilon_i,
\end{align}
where $\mu_j = E(X_j)$, $m_j = E\{\kappa(\cdot,X_j)\}$, $A_{i,j}\in\mathcal{B}_{\mathrm{HS}}(\mathcal{H}_X,\mathcal{H}_X)$, $B_{i,j}\in\mathcal{B}_{\mathrm{HS}}(\mathcal{H}_X,\mathscr{H}_X)$, and $\epsilon_1,\dots,\epsilon_p$ are mutually independent mean-zero random functions in $\mathcal{H}_X$.
\end{defi}
The identifiability condition in \citet{Lee2022} requires purely non-linear relationships, and thus FMANM cannot be used in their methodology. In contrast, the use of FMANM here takes advantage of no structural or distributional requirement of our identifiability condition. FMANM provides a unified framework that subsumes purely linear and purely non-linear models, which may be of independent interest in the field of functional regression. 

To estimate a FMANM, in addition to the covariance operator $\Gamma_{X_iX_j} = E\{(X_i-\mu_i)\otimes (X_j-\mu_j)\}$, we further define the cross-covariance operator $\Lambda_{X_iX_j}: \mathcal{H}_{X} \to \mathscr{H}_{X}$, and the kernel-covariance operator $\Sigma_{X_iX_j}: \mathscr{H}_{X} \to \mathscr{H}_{X}$ by, respectively, 
\begin{align*}
\Lambda_{X_iX_j} = E[(X_i-\mu_i)\otimes \{\kappa(\cdot,X_j)-m_j\}],~
\Sigma_{X_iX_j} = E[\{\kappa(\cdot,X_i)-m_i\}\otimes \{\kappa(\cdot,X_j)-m_j\}].
\end{align*}
The three covariance operators above capture different aspects of dependence between functional variables. For example, the cross‑covariance operator $\Lambda_{X_iX_j}: \mathcal{H}_{X} \to \mathscr{H}_{X}$ measures the covariance between the original variable $X_i$ and the non-linearly mapped feature $\kappa(\cdot, X_j)$. 

For compact notations, we introduce vector and matrix versions of the operators. Let $\mathbb{S}$ be a subset of $\mathbb{V}$. Define the Hilbert direct sums $\mathcal{H}_{\mathbf{X}_{\mathbb{S}}}=\bigoplus_{j\in\mathbb{S}}\mathcal{H}_{X}$ and $\mathscr{H}_{\mathbf{X}_{\mathbb{S}}}=\bigoplus_{j\in\mathbb{S}}\mathscr{H}_{X}$, whose elements are tuples $(f_j)_{j\in\mathbb{S}}$ with $f_j\in\mathcal{H}_{X}$ and $(\phi_j)_{j\in\mathbb{S}}$ with $\phi_j\in\mathscr{H}_{X}$, respectively, equipped with the inner products $\langle (f_j)_{j\in\mathbb{S}},(g_j)_{j\in\mathbb{S}}\rangle_{\mathcal{H}_{\mathbf{X}_{\mathbb{S}}}}=\sum_{j\in\mathbb{S}}\langle f_j,g_j\rangle_{\mathcal{H}_{X}}$ and $\langle (\phi_j)_{j\in\mathbb{S}},(\psi_j)_{j\in\mathbb{S}}\rangle_{\mathscr{H}_{\mathbf{X}_{\mathbb{S}}}}=\sum_{j\in\mathbb{S}}\langle\phi_j,\psi_j\rangle_{\mathscr{H}_{X}}$. For $\mathbb{S}\subseteq\mathbb{V}$ and $j\in\mathbb{V}$, the block operator $\Gamma_{\mathbf{X}_\mathbb{S} X_j}: \mathcal{H}_{\mathbf{X}_\mathbb{S}} \to \mathcal{H}_{X}$ is defined by its action $\Gamma_{\mathbf{X}_\mathbb{S} X_j}\mathbf{f}=\sum_{i\in\mathbb{S}}\Gamma_{X_iX_j}f_i$ for $\mathbf{f}=(f_i)_{i\in\mathbb{S}}\in\mathcal{H}_{\mathbf{X}_\mathbb{S}}$. More generally, for $\mathbb{S}_k, \mathbb{S}_\ell \subseteq \mathbb{V}$, the block operator matrix $\Gamma_{\mathbf{X}_{\mathbb{S}_k} \mathbf{X}_{\mathbb{S}_\ell}}: \mathcal{H}_{\mathbf{X}_{\mathbb{S}_k}} \to \mathcal{H}_{\mathbf{X}_{\mathbb{S}_\ell}}$ is defined by its action $[\Gamma_{\mathbf{X}_{\mathbb{S}_k} \mathbf{X}_{\mathbb{S}_\ell}}\mathbf{f}]_j=\sum_{i\in\mathbb{S}_k}\Gamma_{X_iX_j}f_i$, for $j\in\mathbb{S}_\ell$ and $\mathbf{f}=(f_i)_{i\in\mathbb{S}_k}\in\mathcal{H}_{\mathbf{X}_{\mathbb{S}_k}}$. The corresponding block operators for $\Lambda$ and $\Sigma$ are defined analogously. 

The following unified covariance operator $\mathcal{V}_\mathbb{S}$ on $\mathcal{H}_{\mathbf{X}_\mathbb{S}}\times\mathscr{H}_{\mathbf{X}_\mathbb{S}}$ plays a vital role in FMANM. Letting $\mathbf{Z}_{\mathbb{S}}=\big(\mathbf{X}_{\mathbb{S}},\{\kappa(\cdot,X_j)\}_{j\in \mathbb{S}}\big)^\top\in \mathcal{H}_{\mathbf{X}_\mathbb{S}}\times\mathscr{H}_{\mathbf{X}_\mathbb{S}}$ and $\boldsymbol{\mu}_{\mathbf{Z}_\mathbb{S}}=(\mu_\mathbb{S},m_\mathbb{S})^\top$ with $\mu_\mathbb{S}=(\mu_j)_{j\in\mathbb{S}}$ and $m_\mathbb{S}=(m_j)_{j\in\mathbb{S}}$, we define $\mathcal{V}_\mathbb{S}=E\{(\mathbf{Z}_{\mathbb{S}}-\boldsymbol{\mu}_{\mathbf{Z}_\mathbb{S}})\otimes(\mathbf{Z}_{\mathbb{S}}-\boldsymbol{\mu}_{\mathbf{Z}_\mathbb{S}})\}$, where the tensor product on $\mathcal{H}_{\mathbf{X}_\mathbb{S}}\times\mathscr{H}_{\mathbf{X}_\mathbb{S}}$ is given by $(\mathbf{z}\otimes\mathbf{z}')\mathbf{f}=\langle\mathbf{z},\mathbf{f}\rangle\mathbf{z}'$ with $\langle(\mathbf{f},\boldsymbol{\phi}),(\mathbf{g},\boldsymbol{\psi})\rangle=\langle\mathbf{f},\mathbf{g}\rangle_{\mathcal{H}_{\mathbf{X}_\mathbb{S}}}+\langle\boldsymbol{\phi},\boldsymbol{\psi}\rangle_{\mathscr{H}_{\mathbf{X}_\mathbb{S}}}$. It admits the block representation
\begin{align}\label{def_C}
\mathcal{V}_\mathbb{S}=\begin{pmatrix}\Gamma_{\mathbf{X}_\mathbb{S}\mathbf{X}_\mathbb{S}} & \Lambda_{\mathbf{X}_\mathbb{S}\mathbf{X}_\mathbb{S}}^* \\ \Lambda_{\mathbf{X}_\mathbb{S}\mathbf{X}_\mathbb{S}} & \Sigma_{\mathbf{X}_\mathbb{S}\mathbf{X}_\mathbb{S}}\end{pmatrix},
\end{align}
that is, for any $(\mathbf{f},\boldsymbol{\phi})^\top\in\mathcal{H}_{\mathbf{X}_\mathbb{S}}\times\mathscr{H}_{\mathbf{X}_\mathbb{S}}$,
\begin{align*}
\mathcal{V}_\mathbb{S}\begin{pmatrix}\mathbf{f}\\\boldsymbol{\phi}\end{pmatrix}
=\begin{pmatrix}\Gamma_{\mathbf{X}_\mathbb{S}\mathbf{X}_\mathbb{S}}\mathbf{f}+\Lambda_{\mathbf{X}_\mathbb{S}\mathbf{X}_\mathbb{S}}^*\boldsymbol{\phi}\\ \Lambda_{\mathbf{X}_\mathbb{S}\mathbf{X}_\mathbb{S}}\mathbf{f}+\Sigma_{\mathbf{X}_\mathbb{S}\mathbf{X}_\mathbb{S}}\boldsymbol{\phi}\end{pmatrix}.
\end{align*} 

Next, we introduce several regularity assumptions that guide a consistent estimation procedure.
 
\begin{ass}\label{ass:1} The kernel $\kappa$ satisfies
$|\kappa(f,g)|\le 1$ for all $f,g\in\mathcal{H}_X$, and for any $i\in\mathbb{V}$ and any $\mathbb{S}\subseteq\mathbb{V}\setminus\{i\}$, 
\begin{align*}
\ker(\mathcal{V}_\mathbb{S}) =\{0\}\,,~~\operatorname{ran}\Big\{\Big(\begin{smallmatrix}\Gamma_{X_i\mathbf{X}_\mathbb{S}} \\ \Lambda_{X_i\mathbf{X}_\mathbb{S}}\end{smallmatrix}\Big)\Big\} \subseteq \operatorname{ran}(\mathcal{V}_\mathbb{S}).
\end{align*}
\end{ass}

\begin{ass}\label{ass:2}
For any node $i\in\mathbb{V}$ and any subset $\mathbb{S}\subseteq\mathbb{V}\setminus\{i\}$, there exists a Hilbert--Schmidt operator $\mathbf{R}_{X_i\mathbf{X}_\mathbb S} \in \mathcal{B}_{\mathrm{HS}}\bigl(\mathcal{H}_X,\; \mathcal{H}_{\mathbf{X}_{\mathbb{S}}}\times\mathscr{H}_{\mathbf{X}_{\mathbb{S}}}\bigr)$ such that $E(X_i\mid\mathbf{X}_\mathbb{S}) = \mu_i + \mathbf{R}_{X_i\mathbf{X}_\mathbb{S}}^* (\mathbf{Z}_\mathbb{S} - \boldsymbol{\mu}_{\mathbf{Z}_\mathbb{S}})$.   
\end{ass} 

In Assumption~\ref{ass:1}, the boundedness requirement is mild, while the compatibility requirement on $\mathcal{V}_{\mathbb{S}}$ leads to a closed-form of the regression operator in Proposition~\ref{prop1} below. Assumption~\ref{ass:2} holds automatically for $\mathrm{pa}(i;\mathbb{G}) \subseteq \mathbb{S} \subseteq \mathrm{nd}(i;\mathbb{G})$, since
\begin{align*}
E(X_i \mid \mathbf X_{\mathbb{S}}) = E(X_i \mid \mathbf X_{\mathrm{pa}(i)}) = \mu_i + \sum_{j \in \mathrm{pa}(i;\mathbb G)} \bigl\{ A_{i,j}^* (X_j - \mu_j) + B_{i,j}^* (\kappa(\cdot, X_j) - m_j) \bigr\},
\end{align*}
while Assumption~\ref{ass:2} requires that such a form holds for any subset $\mathbb{S}\subseteq\mathbb{V}\setminus\{i\}$. We refer to $\mathbf{R}_{X_i\mathbf X_\mathbb{S}}$ or its adjoint as a regression operator. Assumptions similar to Assumptions~\ref{ass:1} and \ref{ass:2} have been postulated in \citet{Lee2022} and \citet{Lee2024}.
 
Next, we provide several propositions forming the basis of FMANM, whose proofs are given in Section D of the Supplementary Material.
\begin{prop}\label{prop1}
Under Assumptions \ref{ass:1} and \ref{ass:2}, for $i\in\mathbb V$ and any subset $\mathbb S\subseteq\mathbb V\setminus\{i\}$, we have
\begin{align*}
\Gamma_{X_iX_i\mid\mathbf{X}_\mathbb{S}} = \Gamma_{X_iX_i} - 
\bigl( \Gamma_{\mathbf{X}_\mathbb{S}X_i},\; \Lambda^*_{X_i\mathbf{X}_\mathbb{S}} \bigr)\mathcal{V}_{\mathbb{S}}^{\dagger}
\Big(\begin{smallmatrix}
\Gamma_{X_i\mathbf{X}_\mathbb{S}} \\
\Lambda_{X_i\mathbf{X}_\mathbb{S}}
\end{smallmatrix}\Big), ~~
\mathbf R_{X_i\mathbf{X}_\mathbb{S}} = 
\mathcal{V}_{\mathbb{S}}^{\dagger}
\Big(\begin{smallmatrix}
\Gamma_{X_i\mathbf{X}_\mathbb{S}} \\
\Lambda_{X_i\mathbf{X}_\mathbb{S}}
\end{smallmatrix}\Big),
\end{align*} 
where $\dagger$ denotes the Moore–Penrose inverse \citep{Li2018}.
\end{prop}

The proof of Proposition~\ref{prop1} is similar to that of proposition~2 in \citet{Lee2022}, except that the covariance structure is more involved in our setting. The expressions above lead to plug-in estimators of the conditional covariance and regression operators based on estimators of the unconditional covariance operators. 

\begin{prop}\label{prop2}
Under Assumptions \ref{ass:1} and \ref{ass:2}, for any $\mathbb{S} \subseteq \mathrm{nd}(i;\mathbb G)\setminus\mathrm{pa}(i;\mathbb{G})$, we have $\mathbf R_{X_i\mathbf{X}_{\mathrm{pa}(i;\mathbb{G})\cup\mathbb{S}}}=(\mathbf{C}_{\mathrm{pa}(i;\mathbb{G}),i}, \mathbf 0)$,
where $\mathbf{C}_{\mathrm{pa}(i;\mathbb{G}),i}=(\mathbf C_{j,i})_{j \in \mathrm{pa}(i;\mathbb{G})}=\{(A_{i,j},B_{i,j})^\top\}_{j \in \mathrm{pa}(i;\mathbb{G})}$ with $A_{i,j}$ and $B_{i,j}$ given in \eqref{eq_FMANM}.
Here, the notation $(\mathbf{C}_{\mathrm{pa}(i;\mathbb{G}),i}, \mathbf 0)$ represents that the operator has non-zero components only for the parent indices of $i$ in $\mathbb{G}$, and the zero components correspond to $j\in\mathbb{S}$. 
\end{prop}
From Proposition~\ref{prop2}, we see that when performing regression of the $i^{th}$ function on its parent functions and non-descendant functions, the regression operator is solely determined by its parents. In addition, recall that the error terms $\epsilon_1, \dots, \epsilon_p$ are mutually independent in FMANM, and thus it is a Markovian causal model, i.e., each variable $X_i$ is conditionally independent of all its non-descendants given its parents $\mathbf{X}_{\mathrm{pa}(i;\mathbb{G})}$ \citep{pearl2009}. This motivates us to perform a multivariate function-on-function regression to estimate $\mathbf{C}_{\{\mathrm{pa}(i;\mathbb{G})\},i}$. 

For all $i\in\mathbb V$ and $\mathbb{S}\subseteq\mathbb V\setminus\{i\}$, we consider the objective function of $\mathbf C_{\mathbb{S},i}:\mathcal{H}_X\to\mathcal{H}_{\mathbf{X}_\mathbb{S}}\times\mathscr{H}_{\mathbf{X}_\mathbb{S}}$,
\begin{align*}
\mathcal{L}_i(\mathbf C_{\mathbb{S},i}) = E\Bigl[ \operatorname{tr}\Bigl\{ \Bigl( X_i - \mu_i - \sum_{j\in\mathbb{S}} \mathbf C_{j,i}^* (\mathbf Z_j - \boldsymbol{\mu}_{\mathbf Z_j}) \Bigr) \otimes \Bigl( X_i - \mu_i - \sum_{j\in\mathbb{S}} \mathbf C_{j,i}^* (\mathbf Z_j - \boldsymbol{\mu}_{\mathbf Z_j}) \Bigr) \Bigr\} \Bigr].
\end{align*}
This objective function designed for FMANM is an extension of the classical least-squares type function to our context. It unifies linear and non-linear components that distinguish it from the related literature \citep{Cai2022,Lee2022}.

\begin{prop}\label{prop3}
If Assumptions \ref{ass:1} and \ref{ass:2} hold, then for any $ i \in \mathbb V $ and $ \mathbb{S} \subseteq \mathbb V \setminus \{i\} $, there exists a constant $ c \ge 0 $ independent of $ \mathbf C_{\mathbb{S},i} $, such that  
\begin{equation}
\mathcal{L}_i(\mathbf C_{\mathbb{S},i}) = 
\left\langle 
\mathcal{V}_{\mathbb{S}}(
\mathbf C_{\mathbb{S},i}),
\mathbf C_{\mathbb{S},i}
\right\rangle_{\mathrm{HS}}
- 2
\Big\langle \Big(
\begin{smallmatrix}
\Gamma_{X_i\mathbf X_\mathbb{S}} \\
\Lambda_{X_i\mathbf X_\mathbb{S}}
\end{smallmatrix}\Big),
\mathbf C_{\mathbb{S},i}
\Big\rangle_{\mathrm{HS}}
+ c. \label{eq:mixed_obj}
\end{equation}
Moreover, we have
\begin{align*}
\mathbf R_{X_i\mathbf X_\mathbb{S}} = 
\operatorname*{argmin}_{\mathbf C_{\mathbb{S},i} \in \mathcal{B}_{\mathrm{HS}}(\mathcal{H}_{X},\, \mathcal{H}_{\mathbf X_\mathbb{S}}\times\mathscr{H}_{\mathbf X_{\mathbb{S}}})}
\mathcal{L}_i(\mathbf C_{\mathbb{S},i}).
\end{align*}
\end{prop}
Proposition~\ref{prop3} provides a quadratic formulation of $\mathcal{L}_i(\mathbf C_{\mathbb{S},i})$ and shows that its minimizer is the regression operator $\mathbf R_{X_i\mathbf X_\mathbb{S}}$. This generalizes the finite-dimensional least-squares solution to Hilbert spaces, jointly accounting for both linear and non-linear dependencies through the unified covariance structure. In particular, FMANM requires inversion of the unified operator $\mathcal{V}_{\mathbb{S}}$, which captures the coupling between the original variables and their kernel embeddings. 

Together, the propositions above provide a rigorous framework for sparse estimation in FMANM, where variable selection corresponds to identifying non-zero blocks of $\mathbf C_{\mathbb{S},i}$, i.e., either the linear component $A_{i,j}$ or the non-linear component $B_{i,j}$ (or both) being non-zero.

\subsection{Estimation}\label{subsc_est}

In this subsection, we establish estimators for FMANM for fully observed functional data. Together with the identifiability results in Section~\ref{sc_fanm}, this yields a complete procedure for identifying a DAG. 

Let $\{(X_{1k}, \dots, X_{pk})\}_{k=1}^n$ be an independent and identically distributed (i.i.d.) sample from the distribution of  $(X_1, \dots, X_p)$. Denote by $E_n$ the empirical mean, i.e., for any random element $\omega$, $E_n(\omega) = \frac{1}{n} \sum_{k=1}^n \omega_k$, where $(\omega_k)_{k=1}^n$ are i.i.d.~copies of $\omega$. 
For any $(i,j) \in \mathbb{V} \times \mathbb{V}$, define the sample mean functions $\hat{\mu}_i = E_n(X_i)$, $\hat{m}_j = E_n\{\kappa(\cdot,X_j)\}$, and the sample covariance operators
\begin{align*}
\hat{\Gamma}_{X_iX_j} &= E_n\{(X_i - \hat{\mu}_i) \otimes (X_j - \hat{\mu}_j)\},~
\hat{\Lambda}_{X_iX_j} = E_n[(X_i - \hat{\mu}_i) \otimes \{\kappa(\cdot,X_j) - \hat{m}_j\}],\\
\hat{\Sigma}_{X_iX_j} &= E_n[\{\kappa(\cdot,X_i) - \hat{m}_i\} \otimes \{\kappa(\cdot,X_j) - \hat{m}_j\}].
\end{align*} 
The vector and matrix estimators can be defined accordingly. To accelerate computation, following \citet{Lee2022}, we further consider the following truncated estimators,
\begin{align*}
\hat{\Gamma}^d_{X_iX_j} &= \sum_{k,\ell=1}^d E_n(\hat{x}_{ik}\hat{x}_{j \ell})(\hat{f}_{ik}\otimes \hat{f}_{j \ell}), ~
\hat{\Lambda}^d_{X_iX_j} = \sum_{k,\ell=1}^d E_n(\hat{x}_{ik}\hat{\alpha}_{j\ell})(\hat{f}_{ik}\otimes \hat{\phi}_{j\ell}), \\
\hat{\Sigma}^d_{X_iX_j} &= \sum_{k,\ell=1}^d E_n(\hat{\alpha}_{ik}\hat{\alpha}_{j\ell})(\hat{\phi}_{ik}\otimes \hat{\phi}_{j\ell}),
\end{align*}
where $\{(\hat{a}_{ik},\hat{f}_{ik})\}_{k=1}^d$ and $\{(\hat{b}_{ik},\hat{\phi}_{ik})\}_{k=1}^d$ are the empirical eigenvalue--eigenfunction pairs of $\hat{\Gamma}_{X_iX_i}$ and $\hat{\Sigma}_{X_iX_i}$, respectively, and the sample coefficients are given by $\hat{x}_{im}=\langle X_i-\hat{\mu}_i,\hat{f}_{im}\rangle_{\mathcal{H}_X}$ and $\hat{\alpha}_{im}=\langle \kappa(\cdot,X_i)-\hat{m}_i,\hat{\phi}_{im}\rangle_{\mathscr{H}_X}$. 

By Proposition~\ref{prop1}, for any $i\in\mathbb V$ and any $\mathbb{S}\subseteq \mathbb V\setminus\{i\}$, we estimate the conditional covariance operator $\Gamma_{X_iX_i|\mathbf{X}_\mathbb{S}}$ by 
\begin{align}
\hat{\Gamma}^{d,\epsilon}_{X_iX_i|\mathbf{X}_\mathbb{S}} = \hat{\Gamma}^d_{X_iX_i} - 
( \hat{\Gamma}^d_{\mathbf{X}_\mathbb{S}X_i},\; (\hat{\Lambda}^d_{X_i\mathbf{X}_\mathbb{S}})^* )
(\hat{\mathcal{V}}^d_{\mathbb{S}}
+ \epsilon  I)^{-1}
\Big(\begin{smallmatrix}
\hat{\Gamma}^d_{X_i\mathbf{X}_\mathbb{S}} \\
\hat{\Lambda}^d_{X_i\mathbf{X}_\mathbb{S}}
\end{smallmatrix}\Big),\label{est_con}
\end{align}
where
\begin{align*}
\hat{\mathcal{V}}^d_{\mathbb{S}}=
\begin{pmatrix}
\hat{\Gamma}^d_{\mathbf{X}_\mathbb{S}\mathbf{X}_\mathbb{S}} & (\hat{\Lambda}^d_{\mathbf{X}_\mathbb{S}\mathbf{X}_\mathbb{S}})^* \\
\hat{\Lambda}_{\mathbf{X}_\mathbb{S}\mathbf{X}_\mathbb{S}}^d & \hat{\Sigma}_{\mathbf{X}_\mathbb{S}\mathbf{X}_\mathbb{S}}^d
\end{pmatrix}.
\end{align*}
Here, $\epsilon>0$ is a ridge parameter and $I$ is the identity mapping. The minimized conditional covariance ordering $\pi^{\min}_{1\to p}$ is estimated by $\hat{\pi}^{\min}_{1\to p}=(\hat{\pi}^{\min}_1,\dots,\hat{\pi}^{\min}_p)$, where with $\hat{\mathbb{S}}_1 = \emptyset$ and $\hat{\mathbb{S}}_{i} = \{\hat{\pi}^{\min}_1,\dots,\hat{\pi}^{\min}_{i-1}\}$,
\begin{align*}
\hat{\pi}^{\min}_i = \arg\min_{j \in \mathbb{V} \setminus \hat{\mathbb{S}}_{i}} \|\hat{\Gamma}^{d,\epsilon}_{X_jX_j|\mathbf{X}_{\hat{\mathbb{S}}_{i}}}\|_{\mathrm{HS}}, \quad i=1,\dots,p.
\end{align*}

Replacing the operators in \eqref{eq:mixed_obj} by their estimators and dropping the constant $c$, we define, for any node $i\in\mathbb V$ and any subset $\mathbb S\subseteq\mathbb V\setminus\{i\}$, the empirical criterion
\begin{align}\label{eq:pop_obj}
\widehat{\mathcal{L}}_{i,0}(\mathbf C_{\mathbb S,i})= \big\langle \hat{\mathcal{V}}^d_{\mathbb S}\mathbf C_{\mathbb S,i},
\mathbf C_{\mathbb S,i}
\big\rangle_{\mathrm{HS}}- 2\Big\langle 
\Big(\begin{smallmatrix}
\hat{\Gamma}^d_{X_i\mathbf{X}_{\mathbb S}} \\
\hat{\Lambda}^d_{X_i\mathbf{X}_{\mathbb S}}
\end{smallmatrix}\Big), \mathbf C_{\mathbb S,i}
\Big\rangle_{\mathrm{HS}}.
\end{align}
To achieve variable selection, we further introduce a penalty function to encourage the sparsity of $\mathbf C_{\mathbb S,i}$. Let $\mathbf{1}\{\cdot\}$ denote the indicator function and $(x)_+ = \max(x,0)$ denote the positive part of $x$. We employ the SCAD penalty proposed by \citet{Fan2001}, whose derivative is defined as
\begin{align}
J'_{\lambda}(\theta) = \lambda\Big\{\mathbf{1}\{\theta\leq\lambda\} + \frac{(a\lambda-\theta)_+}{(a-1)\lambda} \mathbf{1}\{\theta>\lambda\}\Big\},\label{SCAD}
\end{align}
for $a>2$, $\theta >0$ and $\lambda>0$ with $J_\lambda(0)=0$, where $\lambda$ is a sparsity tuning parameter controlling the sparsity. The SCAD penalty can produce sparse solutions while simultaneously providing almost unbiased estimates for large coefficients \citep{Fan2001}. Here we set $a = 6$ that is slightly larger than the common choice $a=3.7$ to achieve weaker nonconvexity. The penalized objective function is then defined as
\begin{align}\label{eq:reg_obj}
\hat{\mathcal{L}}_i(\mathbf C_{\mathbb S,i})=\widehat{\mathcal{L}}_{i,0}(\mathbf C_{\mathbb S,i})+\sum_{j\in\mathbb S}J_{\lambda}\big(\|\mathbf C_{j,i}\|_{\mathrm{HS}}\big).
\end{align}
At each ordering stage $r=2,\dots,p$, the regression is obtained by instantiating \eqref{eq:reg_obj} at the response node $i=\hat\pi^{\min}_r$ and the predecessor set $\mathbb S=\hat{\mathbb S}_r=\{\hat\pi^{\min}_1,\dots,\hat\pi^{\min}_{r-1}\}$. 

Let $\hat{\mathcal{H}}^{d}_{X_j}$ and $\hat{\mathscr{H}}^{d}_{X_j}$ denote the subspaces spanned by the first $d$ empirical eigenfunctions $\hat f_{j1},\dots,\hat f_{jd}$ of $\hat\Gamma_{X_jX_j}$ and $\hat\phi_{j1},\dots,\hat\phi_{jd}$ of $\hat\Sigma_{X_jX_j}$, respectively, and let $\hat{\mathcal P}^d_j$ and $\hat{\mathcal Q}^d_j$ denote the orthogonal projections onto them. Let $\hat{\mathbf C}_{\mathbb S,i}$ denote a local minimizer of $\hat{\mathcal L}_i(\mathbf C_{\mathbb S,i})$ over the class of operators $\mathbf C_{\mathbb S,i}=(\mathbf C_{j,i})_{j\in\mathbb S}$, where $\mathbf C_{j,i}=(A_{i,j},B_{i,j})^\top$ satisfies
$A_{i,j}=\hat{\mathcal P}_j^d A_{i,j}\hat{\mathcal P}_i^d$ and $B_{i,j}=\hat{\mathcal Q}_j^d B_{i,j}\hat{\mathcal P}_i^d$, for $j\in\mathbb S$. Here, $\hat{\mathcal P}_j^d$ and $\hat{\mathcal Q}_j^d$ denote
the orthogonal projections onto the spans of the first $d$
eigenfunctions of $\hat\Gamma_{X_jX_j}$ and
$\hat\Sigma_{X_jX_j}$, respectively. This class is thus a finite-dimensional space, and the minimization is aligned with the matrix computation described in Section~\ref{sc_comp}. Since the SCAD-penalized criterion \eqref{eq:reg_obj} is nonconvex, such a minimizer is not necessarily unique. Theorem~\ref{theorem3} establishes the existence of a local minimizer satisfying estimation and
selection consistency. We then estimate the edges of the corresponding DAG for FMANM by
\begin{align*}
\hat{\mathbb{E}}=\left\{(\hat{\pi}^{\min}_j,\hat{\pi}^{\min}_i): \|\hat{\mathbf C}_{\hat{\pi}^{\min}_j,\hat{\pi}^{\min}_i}\|_\mathrm{HS}\neq0,\, i=j+1,\dots,p,\, j=1,\dots,p-1\right\}.
\end{align*} 

\section{Asymptotic Theory}\label{sc_theo} 

We establish the asymptotic uniform consistency of order determination and the convergence rates of regression coefficients in FMANM, which induces the consistency of the two-stage procedure of identifying the DAG. We treat the number of nodes $p$ as fixed, because the mathematical induction in the proof of Theorem~\ref{theorem1} requires a finite $p$ and many additional conditions are required for a diverging $p$. Nevertheless, we retain $p$ in all convergence statements rather than absorbing it into constants, and thus the effect of the graph size on the rates remains transparent. Let $\mathbb{G}^0$ and $\mathbb E^0$ denote the true DAG and edge set, respectively. 
\begin{ass}\label{ass:3}
There exists a positive constant $\tau_F$ such that for every ancestral subset $\mathbb S \subset \mathbb{V}$ and any node $j \in \mathbb{V} \setminus \mathbb S$ with $\mathrm{pa}(j;\mathbb G^0) \subseteq \mathbb S$ and any descendant $k \in \mathrm{de}(j)$, we have
\begin{align*}
\|\Gamma_{X_j X_j \mid \mathbf{X}_\mathbb S}\|_{\mathrm{HS}} + \tau_F < \|\Gamma_{X_k X_k \mid \mathbf{X}_\mathbb S}\|_{\mathrm{HS}}.
\end{align*}
\end{ass}
\begin{ass}\label{ass:4}
There exists $\beta>0$ such that $\max_{i\in \mathbb V}(\sum^\infty_{k=d+1}a_{ik})\preceq d^{-2\beta}$ and $\max_{i\in \mathbb V}(\sum^\infty_{k=d+1}b_{ik})\preceq d^{-2\beta}$.
\end{ass} 
Assumption~\ref{ass:3} makes the gaps of the conditional covariance norm explicit, which is equivalent to \eqref{eq:2.2} in Theorem~\ref{theorem1} under fixed $p$ and distribution. This is used to derive consistency accounting for estimation error.  Assumption~\ref{ass:4} requires the tail sum of eigenvalues for all functional variables to decay at a polynomial rate, which is commonly postulated in the literature~\citep{Lee2022,Cai2022,Lee2024}. 

Recall that $\hat{\Gamma}^{d,\epsilon}_{X_iX_i|\mathbf{X}_{\mathbb S}}$ is the $d$-truncated ridge-regularized estimator of the conditional covariance operator in \eqref{est_con}. We require that $\nu_d=\min\{\min(|a_{ik}-a_{i\ell}|,|b_{ik}-b_{i\ell}|):1\leq k<\ell\leq d+1,i\in \mathbb V\}>0$, where $|a_{ik}-a_{i\ell}|$ and $|b_{ik}-b_{i\ell}|$ are the distances among all $d+1$ leading eigenvalues of $\Gamma_{X_iX_i}$ and $\Sigma_{X_iX_i}$ for all $i\in \mathbb V$. Recall that $\pi^{\min}_{1\to p} = (\pi^{\min}_1, \dots, \pi^{\min}_p)$ is the minimized conditional covariance ordering based on population quantities. According to Theorem~\ref{theorem1}, we have $\pi^{\min}_{1\to p} \in \Pi(\mathbb G^0)$.
\begin{theo}\label{theorem2}
Under Assumptions \ref{ass:1} to \ref{ass:4}, we have
$$\max_{\mathbb S\subsetneq\mathbb V}\max_{i\in\mathbb V\setminus\mathbb S}\|\hat{\Gamma}^{d,\epsilon}_{X_iX_i|\mathbf{X}_{\mathbb S}}-\Gamma_{X_iX_i|\mathbf{X}_{\mathbb S}}\|_{\mathrm{HS}}=O_p(\mathcal{E}_{n,1}),$$
where $\mathcal{E}_{n,1}=d p^2/(\sqrt{n} \nu_d\epsilon)+\epsilon^{-1}pd^{-\beta}+\epsilon$. If $\mathcal{E}_{n,1}\to 0$, then $P(\hat{\pi}^{\min}_{1\to p}\in\Pi(\mathbb G^0))\to 1$ as $n\to\infty$.
\end{theo} 
Theorem~\ref{theorem2} establishes the consistency of the causal order estimation for FMANM: it derives the convergence rate in probability of $\max_{\mathbb S,i}\|\hat{\Gamma}^{d,\epsilon}_{X_iX_i|\mathbf{X}_{\mathbb S}}-\Gamma_{X_iX_i|\mathbf{X}_{\mathbb S}}\|_{\mathrm{HS}}$, under which the estimated ordering $\hat{\pi}^{\min}_{1\to p}$ is a valid topological ordering of $\mathbb G^0$ with probability tending to one. Since the greedy procedure in \eqref{eq:2.1} may encounter ties, we do not single out a particular population ordering that $\hat{\pi}^{\min}_{1\to p}$ is converging to. 

\begin{ass}\label{ass:5}
There exists a positive sequence $\{c_d\}_{d\ge1}$ such that, for any nonempty subsets $\mathbb{S}\subseteq \mathbb V$ and every $d\ge1$, $\lambda_{\min}(\mathcal{V}^d_{\mathbb{S}}\bigr|_{\mathcal{M}^d_{\mathbb{S}}})\ge c_d$, where $\mathcal{M}^d_{\mathbb{S}}$ denotes the $2d|\mathbb{S}|$-dimensional subspace of $\mathcal{H}_{\mathbf{X}_{\mathbb{S}}}\times\mathscr{H}_{\mathbf{X}_{\mathbb{S}}}$ spanned by the first $d$ eigenfunctions of $\Gamma_{X_jX_j}$ and $\Sigma_{X_jX_j}$ for $j\in\mathbb{S}$, and $\lambda_{\min}(\cdot)$ denotes the minimum eigenvalue. In addition, there exists a constant $s>0$ such that 
\begin{align*}
\max_{i\in\mathbb V}\max_{j\in\mathrm{pa}(i;\mathbb G^0)}&\bigl\{\bigl\|(\mathcal I-\mathcal P^d_j)A_{i,j}\bigr\|_{\mathrm{HS}}+\bigl\|(\mathcal I-\mathcal Q^d_j)B_{i,j}\bigr\|_{\mathrm{HS}}\\
&\qquad+\bigl\|A_{i,j}(\mathcal I-\mathcal P^d_i)\bigr\|_{\mathrm{HS}}+\bigl\|B_{i,j}(\mathcal I-\mathcal P^d_i)\bigr\|_{\mathrm{HS}}\bigr\}\preceq d^{-s},
\end{align*}
where $\mathcal P^d_j$ and $\mathcal Q^d_j$ denote the projections onto the subspaces spanned by the first $d$ eigenfunctions of $\Gamma_{X_jX_j}$ and $\Sigma_{X_jX_j}$, respectively. Finally, the tuning parameters satisfy $d^{-s}\sqrt{p}=o(c_d)$. 
\end{ass} 
The first part of Assumption~\ref{ass:5} requires the truncated unified covariance operator to be positive definite on the truncation subspace, which guarantees that the truncated regression problem is well-posed. The second part imposes a smoothness condition on the true coefficient operators, and the requirement $d^{-s}\sqrt{p}=o(c_d)$ ensures that the truncation bias of the coefficient operators is negligible relative to $c_d$. Similar eigenvalue and smoothness conditions can be found in \citet{Cai2022}.  

We now establish the consistency of the regression estimation. Let $B_n = \{\hat{\pi}^{\min}_{1\to p} \in \Pi(\mathbb G^0)\}$ denote the event that the estimated ordering is a valid topological ordering of $\mathbb G^0$; by Theorem~\ref{theorem2}, $P(B_n)\to1$ as $n\to\infty$. For any node $i$, let $\hat v_i$ denote the position of node $i$ in $\hat{\pi}^{\min}_{1\to p}$, that is, $\hat{\pi}^{\min}_{\hat v_i} = i$, and let $\hat{\mathbb{S}}_{\hat v_i} = \{\hat{\pi}^{\min}_1, \dots, \hat{\pi}^{\min}_{\hat v_i-1}\}$ denote its estimated predecessor set. Let $\mathbf C^0_{\mathbb S,i}$ denote the population regression coefficient operator of $X_i$ on $\mathbf X_\mathbb S$. By Proposition~\ref{prop1}, we have $\mathbf C^0_{\mathbb S,i}=\mathcal V_\mathbb S^\dagger\mathcal W_\mathbb S$.

\begin{theo}\label{theorem3}
Suppose that Assumptions \ref{ass:1} to \ref{ass:5} hold, $\lambda\to0$, $\lambda^{-1}\min_{(j,i)\in\mathbb E^0}\|\mathbf C^0_{j,i}\|_{\mathrm{HS}}\to\infty$, $\mathcal{E}_{n,1}\to0$ and $p\,\mathcal{E}_{n,2}=o(\lambda)$, where $\mathcal{E}_{n,2} = d p^{5/2}/(\sqrt{n} \nu_d c_d) + d^{-s}p^{3/2}/c_d$. Then, for every node $i\in\mathbb V$ with $\hat v_i\ge2$, we have 
\begin{enumerate}
    \item (Estimation consistency) there exists a local minimizer $\hat{\mathbf C}_{\hat{\mathbb S}_{\hat v_i},i}$ of $\hat{\mathcal L}_{i}(\mathbf C_{\hat{\mathbb S}_{\hat v_i},i})$ such that
    $$\mathbf 1\{B_n\}\,\|\hat{\mathbf C}_{\hat{\mathbb S}_{\hat v_i},i}-\mathbf C^0_{\hat{\mathbb S}_{\hat v_i},i}\|_{\mathrm{HS}} = O_p ( \mathcal{E}_{n,2} );$$ 
    \item (Selection consistency) $P\bigl(B_n\cap\bigl\{\|\hat{\mathbf C}_{j,i}\|_{\mathrm{HS}} = 0 \text{ for all } j \in \hat{\mathbb S}_{\hat v_i}\setminus\mathrm{pa}(i;\mathbb G^0),\ \|\hat{\mathbf C}_{j,i}\|_{\mathrm{HS}}\ge\|\mathbf C^0_{j,i}\|_{\mathrm{HS}}/2>0 \text{ for all } j\in\mathrm{pa}(i;\mathbb G^0)\bigr\}\bigr) \to 1$ as $n\to \infty$,
\end{enumerate}
where $P(B_n)\to 1$ as $n\to\infty$ by Theorem~\ref{theorem2}.
\end{theo} 
The proof of Theorem~\ref{theorem3} is given in Section C of the Supplementary Material. Theorem~\ref{theorem3} establishes both estimation and selection consistency for the SCAD-penalized regression.  \citet{Cai2022} also employed the SCAD penalty for variable selection in the function-on-function linear model. In contrast, our model includes a non-linear component that induces more involved technical arguments. 
 
\begin{cor}\label{cor1}
Supposing that Assumptions \ref{ass:1} to \ref{ass:5} hold, $\mathcal{E}_{n,1}\to0$, $\lambda\to0$, $\lambda^{-1}\min_{(j,i)\in\mathbb E^0}\|\mathbf C^0_{j,i}\|_{\mathrm{HS}}\to\infty$, and $p\mathcal{E}_{n,2}=o(\lambda)$, where $\mathcal{E}_{n,2} = d p^{5/2}/(\sqrt{n} \nu_d c_d) + d^{-s}p^{3/2}/c_d$, we have $P(\mathbb{\hat{E}}\neq \mathbb{E}^0)\to 0$ as $n\to \infty$, where $\hat{\mathbb E}$ is induced by $\hat{\mathbf C}_{\hat{\mathbb S}_{\hat v_i},i}$ in Theorem~\ref{theorem3}.
\end{cor}
The proof of Corollary~\ref{cor1} is given in Section C of the Supplementary Material. Corollary~\ref{cor1} combines the results from Theorems~\ref{theorem2} and~\ref{theorem3}, completing the overall consistency of our two-step procedure of identifying the DAG.

The estimation of the conditional covariance operator plays a pivotal role in establishing Theorems \ref{theorem2} and \ref{theorem3}. Since we do not require Gaussianity in our identifiability condition, we only assume bounded fourth moments for $\mathbf X$, which leads to polynomial tail probabilities in the convergence statements. In contrast, under the stronger assumption that $\mathbf X$ are Gaussian (or sub-Gaussian in the least restrictive setting) as in \citet{Lee2024}, one can achieve exponential tail probabilities, which relaxes the rate conditions in Theorems \ref{theorem2} to \ref{theorem3}.

\section{Practical Computation from Discrete Data}\label{sc_comp}

We implement the procedure using finite-dimensional coordinates of the observed curves. When observation times differ, we first smooth each trajectory and evaluate it on a common grid $t_1,\dots,t_q$ \citep{Ramsay2005}. Write $\mathbf x_{ik}=(X_{ik}(t_1),\dots,X_{ik}(t_q))^\top$. Following the kernel-coordinate approach of \citet{Lee2022} and \citet{Lee2024}, we choose a first-layer kernel $\kappa_T$ with Gram matrix $(K_T)_{s,t}=\kappa_T(t_s,t_t)$. Write $K_T=U_TD_TU_T^\top$ and retain its positive eigenvalues. The reconstructed curve has orthonormal coordinates $D_T^{-1/2}U_T^\top\mathbf x_{ik}$. After centering these coordinates, empirical principal component analysis gives the score matrix $\hat{\mathbf Z}_i^{(d)}\in\mathbb R^{n\times d}$. We take $d=\max_i d_i$, where $d_i$ is the smallest number of components explaining at least $92\%$ of the empirical variance at node $i$; rank-deficient cases are detailed in Section~E of the Supplementary Material.

For the non-linear component, we construct the Gram matrix
$\mathscr K_i(k,\ell)=\kappa_i(\hat{\mathbf Z}_{ik}^{(d)},\hat{\mathbf Z}_{i\ell}^{(d)})$, where $\hat{\mathbf Z}_{ik}^{(d)}$ is the $k$-th row of $\hat{\mathbf Z}_i^{(d)}$ and $\kappa_i$ is the nested kernel $\kappa$ applied to $\hat{\mathbf Z}_{ik}^{(d)}$. Let $\mathscr U_i^{(d)}$ and $\boldsymbol\Lambda_i^{(d)}$ contain the leading eigenvectors and eigenvalues of $Q_n\mathscr K_iQ_n$, where $Q_n=I_n-n^{-1}\mathbf1_n\mathbf1_n^\top$. The corresponding centered kernel-score matrix is
$\hat{\mathbf A}_i^{(d)} =\mathscr U_i^{(d)}(\boldsymbol\Lambda_i^{(d)})^{1/2}$. For an ordered set $\mathbb S=(j_1,\dots,j_s)$, define the mixed design $\hat{\mathbf D}_j=[\hat{\mathbf Z}_j^{(d)},\hat{\mathbf A}_j^{(d)}]$, $\hat{\mathbf D}_{\mathbb S}
=[\hat{\mathbf D}_{j_1},\dots,\hat{\mathbf D}_{j_s}]$, and set $\hat{\mathbf V}_{\mathbb S}=n^{-1}\hat{\mathbf D}_{\mathbb S}^\top\hat{\mathbf D}_{\mathbb S}$ and $\hat{\mathbf W}_{i,\mathbb S}=n^{-1}\hat{\mathbf D}_{\mathbb S}^\top\hat{\mathbf Z}_i^{(d)}$. The coordinate version of the conditional covariance estimator is
\begin{equation}\label{eq:comp_cond_main}
\hat{\mathbf G}_{i\mid\mathbb S}
=\frac1n(\hat{\mathbf Z}_i^{(d)})^\top\hat{\mathbf Z}_i^{(d)}
-\hat{\mathbf W}_{i,\mathbb S}^\top
(\hat{\mathbf V}_{\mathbb S}+\epsilon I)^{-1}
\hat{\mathbf W}_{i,\mathbb S}.
\end{equation}
For $\mathbb S=\emptyset$, the second term is zero. Starting with $\hat{\mathbb S}_1=\emptyset$, we recursively select
\[
\hat\pi_r^{\min}\in
\operatorname*{argmin}_{j\in\mathbb V\setminus\hat{\mathbb S}_r}
\|\hat{\mathbf G}_{j\mid\hat{\mathbb S}_r}\|_{\mathrm F},
\qquad
\hat{\mathbb S}_{r+1}
=\hat{\mathbb S}_r\cup\{\hat\pi_r^{\min}\},
\]
with any fixed deterministic tie-breaking rule. Following \citet{Lee2022}, we use $\epsilon=n^{-2/5}$ as a numerical default, which is a computational heuristic that is not covered by the sufficient rate conditions of Theorem~\ref{theorem2}. 

For each $i=\hat\pi_r^{\min}$ with $r\ge2$, let $\mathbb S=\hat{\mathbb S}_r$. Writing $\mathbf C=\lfloor\mathbf C_{\mathbb S,i}\rfloor$ for its coordinate matrix, the edge selection reduces to
\begin{equation}\label{mixed13}
\operatorname*{min}_{\mathbf C\in\mathbb R^{2d|\mathbb S|\times d}}
\Bigg\{
\frac1n\|\hat{\mathbf Z}_i^{(d)}-\hat{\mathbf D}_{\mathbb S}\mathbf C\|_{\mathrm F}^2
+\sum_{j\in\mathbb S}J_\lambda(\|\mathbf C_j\|_{\mathrm F})
\Bigg\},
\end{equation}
where $\mathbf C_j=\lfloor\mathbf C_{j,i}\rfloor$ is the $2d\times d$ block associated with node $j$. This criterion differs from the operator-level empirical criterion only by a constant within the chosen coordinate representation. We use SCAD with $a=6$, select $\lambda$ by a BIC-type rule, and compute a numerical solution by block proximal-gradient iterations with details given in Section~E of the Supplementary Material. Nonzero fitted blocks define
\begin{equation}\label{mixed14}
\hat{\mathbb E}
=\bigl\{(j,\hat\pi_r^{\min}):
 j\in\hat{\mathbb S}_r,\ 
 \|\lfloor\hat{\mathbf C}_{j,\hat\pi_r^{\min}}\rfloor\|_{\mathrm F}>0,
 \ r=2,\dots,p\bigr\}.
\end{equation}

Section~E of the Supplementary Material provides the coordinate derivations, algorithms, tuning rule, and additional numerical reporting threshold. The implementation uses RKHS reconstructions and data-adaptive truncation as computational surrogates. The continuous-$L^2$ theory in Section~\ref{sc_theo} does not incorporate these choices,  or smoothing and discretization errors.

\section{Numerical Experiments}\label{sc_num}
\subsection{Simulation}\label{sc_sim}
In this section, we conduct simulated experiments to illustrate the finite-sample performance of our methodology compared to existing approaches. In particular, our main approach is the Functional Mixed Additive Noise Model (FMixedANM). To show potential advantage of our mixed regression model, we also consider our minimized conditional covariance ordering combined with a purely linear regression model (i.e., \eqref{eq_FMANM} with $B_{ij}\equiv 0$) and a purely non-linear regression model (i.e., \eqref{eq_FMANM} with $A_{ij}\equiv 0$), referred to as FLinANM and FNonANM, respectively; see Sections~E.1 and E.2 of the Supplementary Material for computational details. In each variant, both the ordering step and the regression step use only the corresponding model. Throughout Section~\ref{sc_num}, the first-layer kernel $\kappa_T$ is the Brownian kernel $\kappa_T(s,t)=\min(s,t)$, $s,t\in[0,1]$. The nested kernel $\kappa_i$ for the non-linear component is the Gaussian kernel $\kappa_i(u,v)=\exp\{-\gamma_f\|u-v\|_{\mathcal{H}_X}^{2}\}$, $u,v\in\mathcal{H}_X$, with the bandwidth $\gamma_f$ selected by the median heuristic and shared across nodes\,\citep{Lee2022}. For competitors, we consider the Functional non-linear Gaussian (FNG) model proposed by \citet{Lee2022} and the Functional Linear Non-Gaussian Bayesian (FLNGB) model proposed by \citet{Zhou2023}. FNG and FLNGB are implemented using their original R code.

Functional data are observed at 100 equally spaced points on $[0,1]$. We consider three graph sizes $p\in\{20,40,60\}$ and two sample sizes $n\in\{200,400\}$ with $100$ Monte Carlo replicates for each scenario. The graphs below under all settings are generated independently, except that in model (iv), the graph is a deterministic binary tree given $p$. Throughout the simulations, candidate edges are restricted to $j\to i$ with $j<i$, and thus $(1,2,\dots,p)$ is always a topological ordering of the true DAG $\mathbb G^0$, where the data are generated sequentially along this ordering once the edges are given. The restriction $j<i$ is used only to generate DAGs; the estimation procedures are not supplied with this ordering.

\begin{figure}[t!]
    \centering
    \vspace{-0.5cm}
    \begin{subfigure}[b]{0.48\textwidth}
        \centering
        \caption*{Structural Hamming distance}
        \includegraphics[width=\textwidth, height=10.5cm]{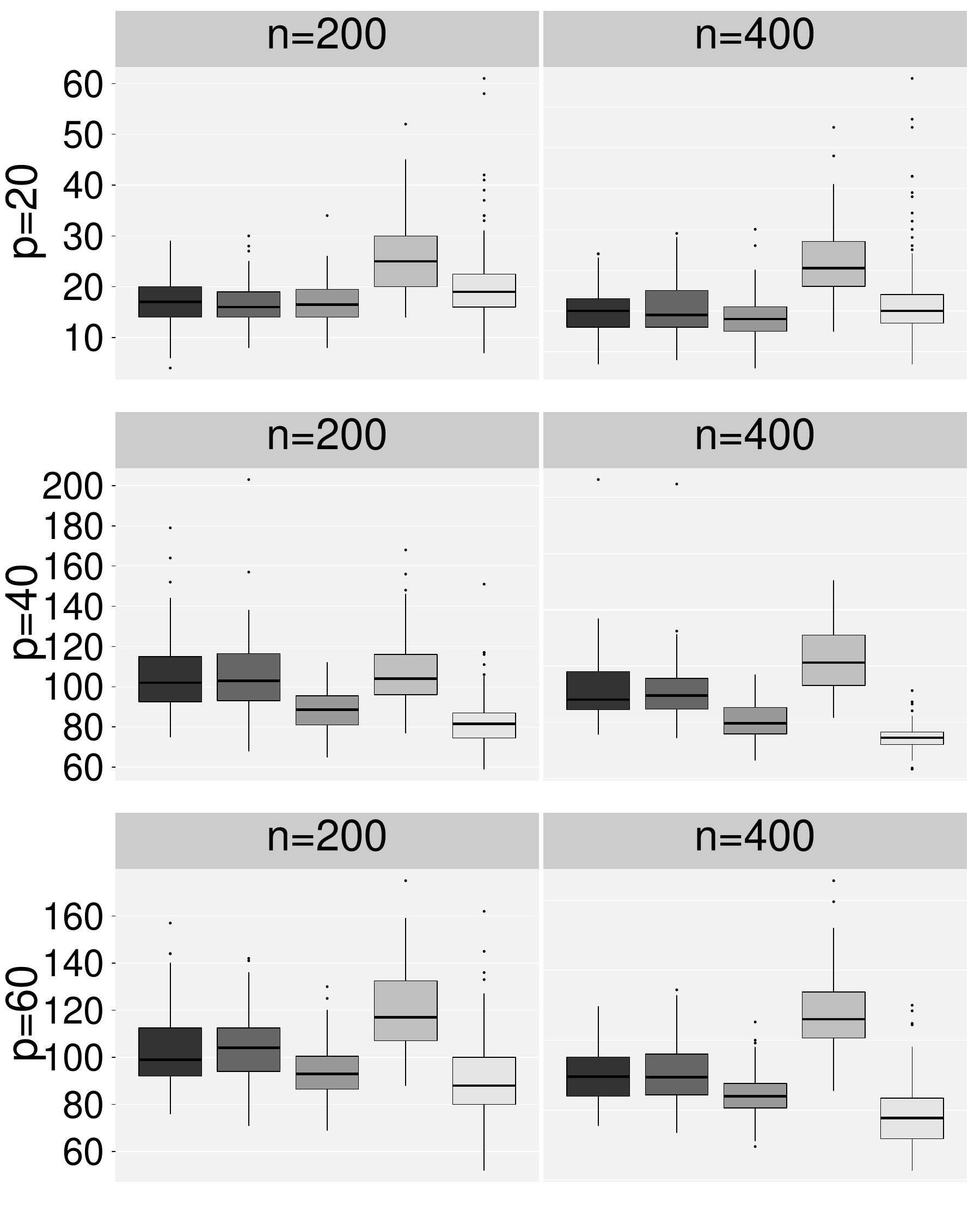}   
        \label{fig:sub1}
    \end{subfigure}
    \hfill 
    \begin{subfigure}[b]{0.48\textwidth}
        \centering
        \caption*{F1 score}
        \includegraphics[width=\textwidth, height=10.5cm]{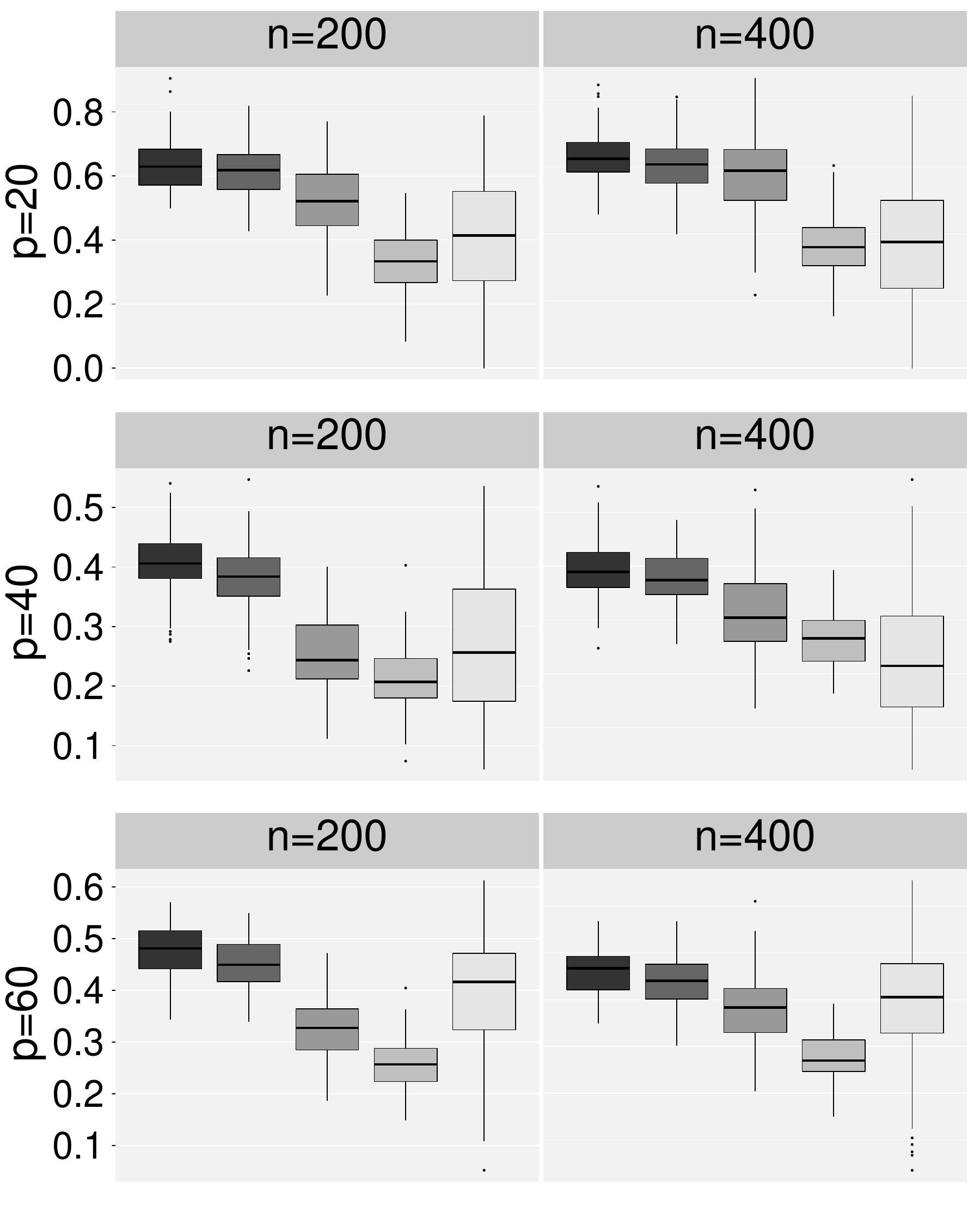}
        \label{fig:sub4}
    \end{subfigure}
    \vspace{-1.2cm}
    \caption{Simulation results under model (i); the boxes from left to right represent the five methods, FMixedANM, FLinANM, FNonANM, FNG, and FLNGB.}
    \label{fig:1} 
\end{figure}

The linear models (i) and (ii) below are similar to those in \citet{Zhou2023}. We obtain the basis $\{\phi_k\}_{k=1}^5$ by orthogonalizing $\phi_k^U(t)=\sum_{\ell=1}^6 A_{k\ell}b_\ell(t)$, where  $b_\ell$ is a cubic B-spline basis function with evenly spaced knots and $A_{k\ell}\sim\mathcal N(0,1)$, for $t\in[0,1]$. We then generate $\mathbf Z_i=(Z_{i1},\dots,Z_{i5})^\top$ and set $X_i(t) =\sum_{j=1}^5 Z_{ij}\phi_j(t)$, where 
\begin{align*}
\mathbf Z_1 = \boldsymbol\varepsilon_1, ~
\mathbf Z_i &= \sum_{j\in\mathrm{pa}(i;\mathbb G^0)}\mathbf B_{ij}\mathbf Z_j + \boldsymbol\varepsilon_i,\qquad i=2,\dots,p. 
\end{align*}
Here, all entries in $\mathbf B_{ij}\in\mathbb R^{5\times 5}$ are drawn from $\mathcal{N}(0,1)$ if there exists an edge $j\to i$, and $\mathbf B_{ij}=\mathbf 0$ otherwise. The edges are generated from independent Bernoulli variables with probability $0.1$ of being an edge for $p=20,40$, and with probability $0.05$ for $p=60$. We consider two noise settings:
\begin{itemize}
\item[(i)] (\emph{Gaussian noise}) $\varepsilon_{ik}\sim\mathcal N(0,0.02)$ ;
\item[(ii)] (\emph{Non-Gaussian noise}) $\varepsilon_{ik}\sim\mathrm{Unif}(-\sqrt{0.06},\sqrt{0.06})$,
\end{itemize}  
where in both settings the $\varepsilon_{ik}$'s are mutually independent and independent of any random generated elements. 

\begin{figure}[t!]
    \centering
    \vspace{-0.5cm}
    \begin{subfigure}[b]{0.48\textwidth}
        \centering
        \caption*{Structural Hamming distance}
        \includegraphics[width=\textwidth, height=10.5cm]{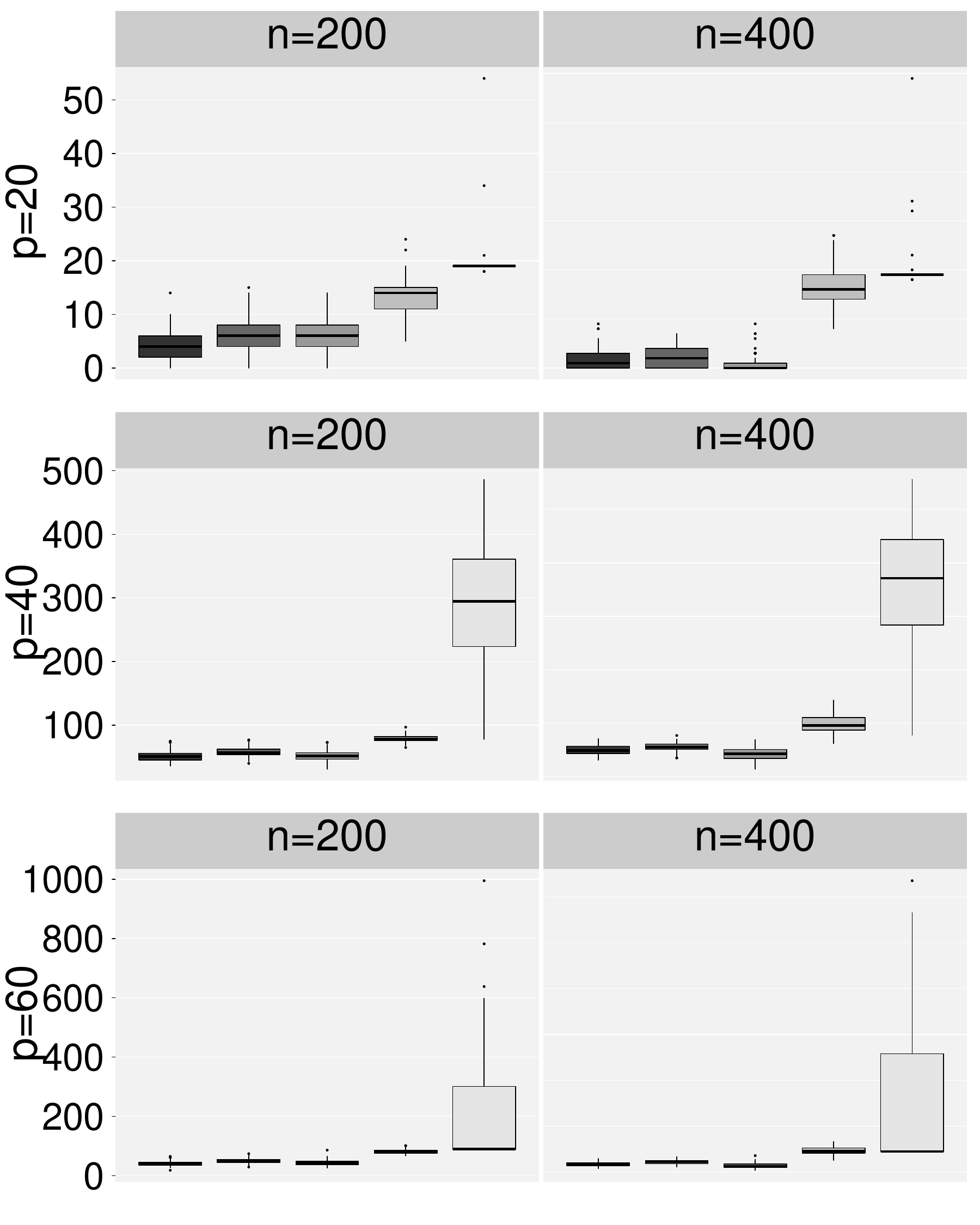}
        \label{fig:sub1_nlg}
    \end{subfigure}
    \hfill 
    \begin{subfigure}[b]{0.48\textwidth}
        \centering
        \caption*{F1 score}
        \includegraphics[width=\textwidth, height=10.5cm]{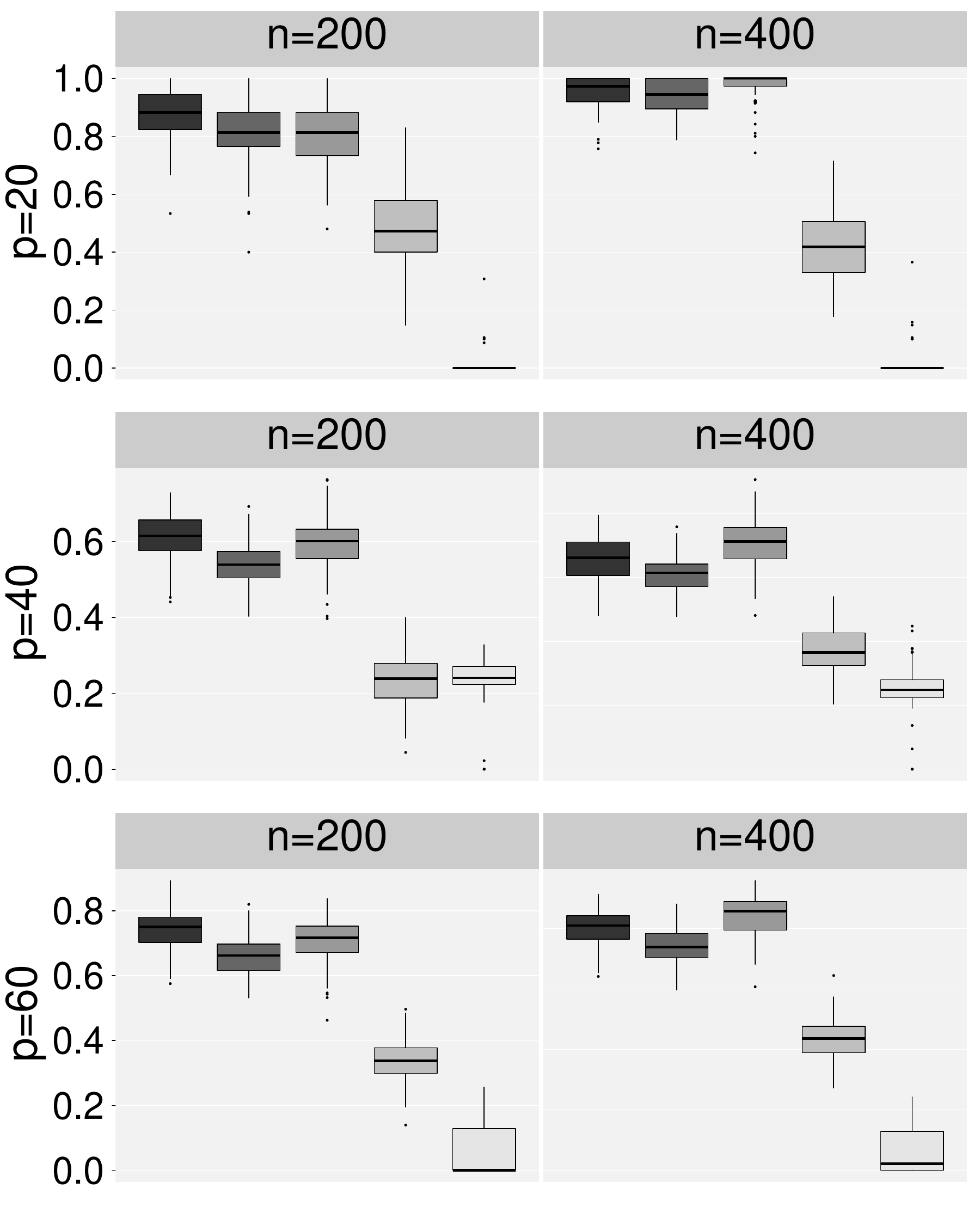}
        \label{fig:sub4_nlg}
    \end{subfigure}
    \vspace{-1.2cm} 
    \caption{Simulation results under model (iii); the boxes from left to right represent the five methods, FMixedANM, FLinANM, FNonANM, FNG, and FLNGB.}
    \label{fig:3} 
\end{figure}

The non-linear models (iii)--(v) below are variants of those in \citet{Lee2022}. We set $X_i(t)=\sum_{k=1}^{20} c_{i,k}\sqrt{2}\sin\{(k-1/2)\pi t\}$ with $J=20$, where
\begin{align*}
c_{1,k} = \xi_{1,k}, ~c_{i,k} = \sum_{j\in\mathrm{pa}(i;\mathbb G^0)} g_k(c_{j,k}) + \xi_{i,k}, \qquad i=2,\dots,p,\; k=1,\dots,J.
\end{align*}
Here, $g_k(x)=x$ if $k$ is even and $g_k(x)=-x^2$ otherwise, and the noise follows one of the following settings: 
\begin{itemize}
\item[(iii)] \emph{(Gaussian noise)} $\xi_{i,k}\sim\mathcal N(0,0.02)$;
\item[(iv)] \emph{(Non-Gaussian noise)} $\xi_{i,k}\sim\mathrm{Unif}(-\sqrt{0.3},\sqrt{0.3})$;
\item[(v)] \emph{(Non-Gaussian and heterogeneous noise)} $\xi_{i,k}\sim\mathrm{Unif}(-\sqrt{0.12},\sqrt{0.12})$ for root nodes and $\xi_{i,k}\sim\mathrm{Unif}(-\sqrt{0.03},\sqrt{0.03})$ for other nodes,
\end{itemize}
where in all three settings the $\xi_{ik}$'s are mutually independent and independent of any random generated elements.  Under models (iii) and (v), the edges are generated from independent Bernoulli variables with probability $0.1$ of being an edge for $p=20,40$, and with probability $0.05$ for $p=60$. Under model (iv), we adopt a tree structure where each node $j$ connects to nodes $2j$ and $2j+1$.  

\begin{figure}[t!]
    \centering
    \vspace{-0.5cm}
    \begin{subfigure}[b]{0.48\textwidth}
        \centering
        \caption*{Structural Hamming distance}
        \includegraphics[width=\textwidth, height=10.5cm]{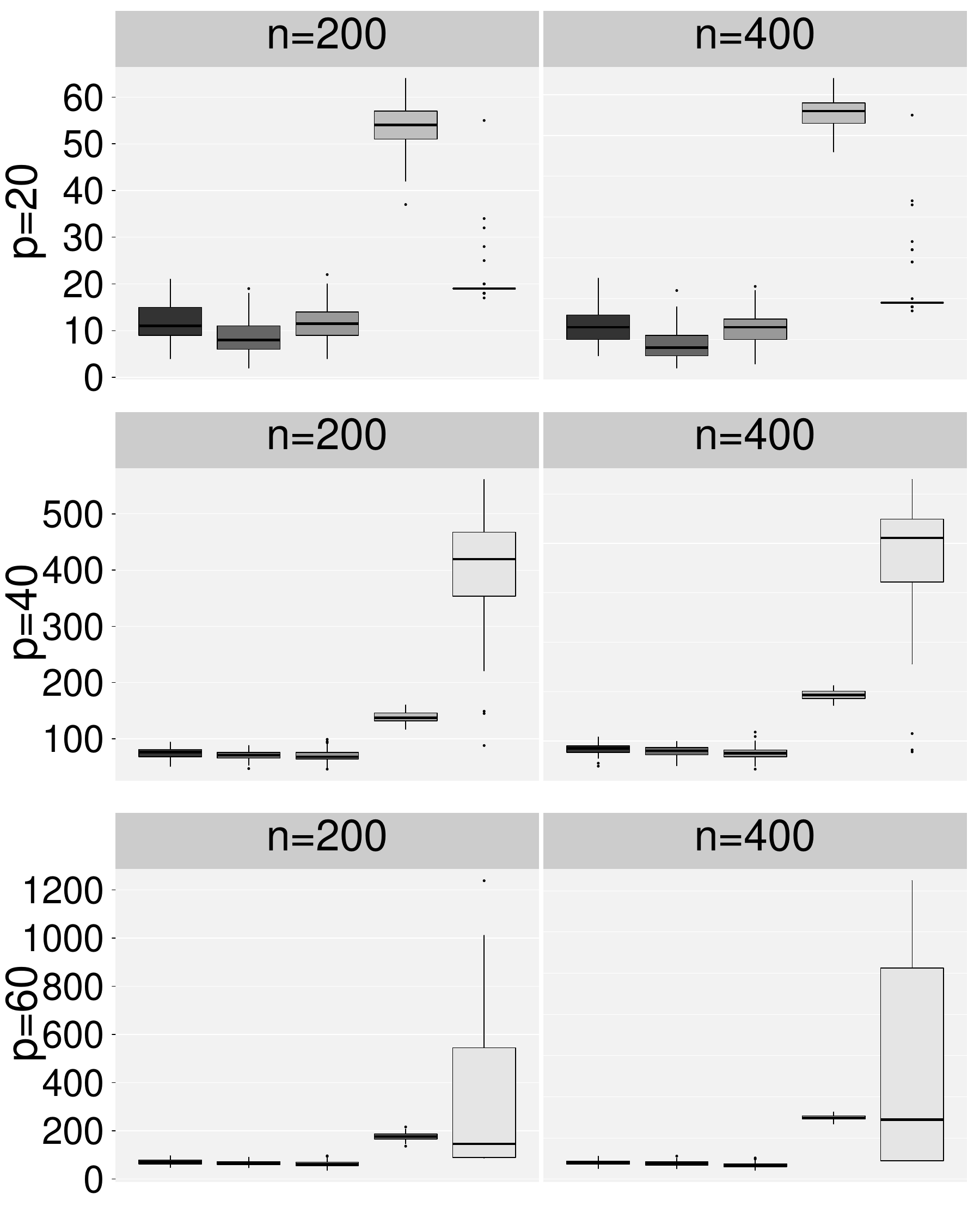}
        \label{fig:sub1_nlngv}
    \end{subfigure}
    \hfill 
    \begin{subfigure}[b]{0.48\textwidth}
        \centering
        \caption*{F1 score}
        \includegraphics[width=\textwidth, height=10.5cm]{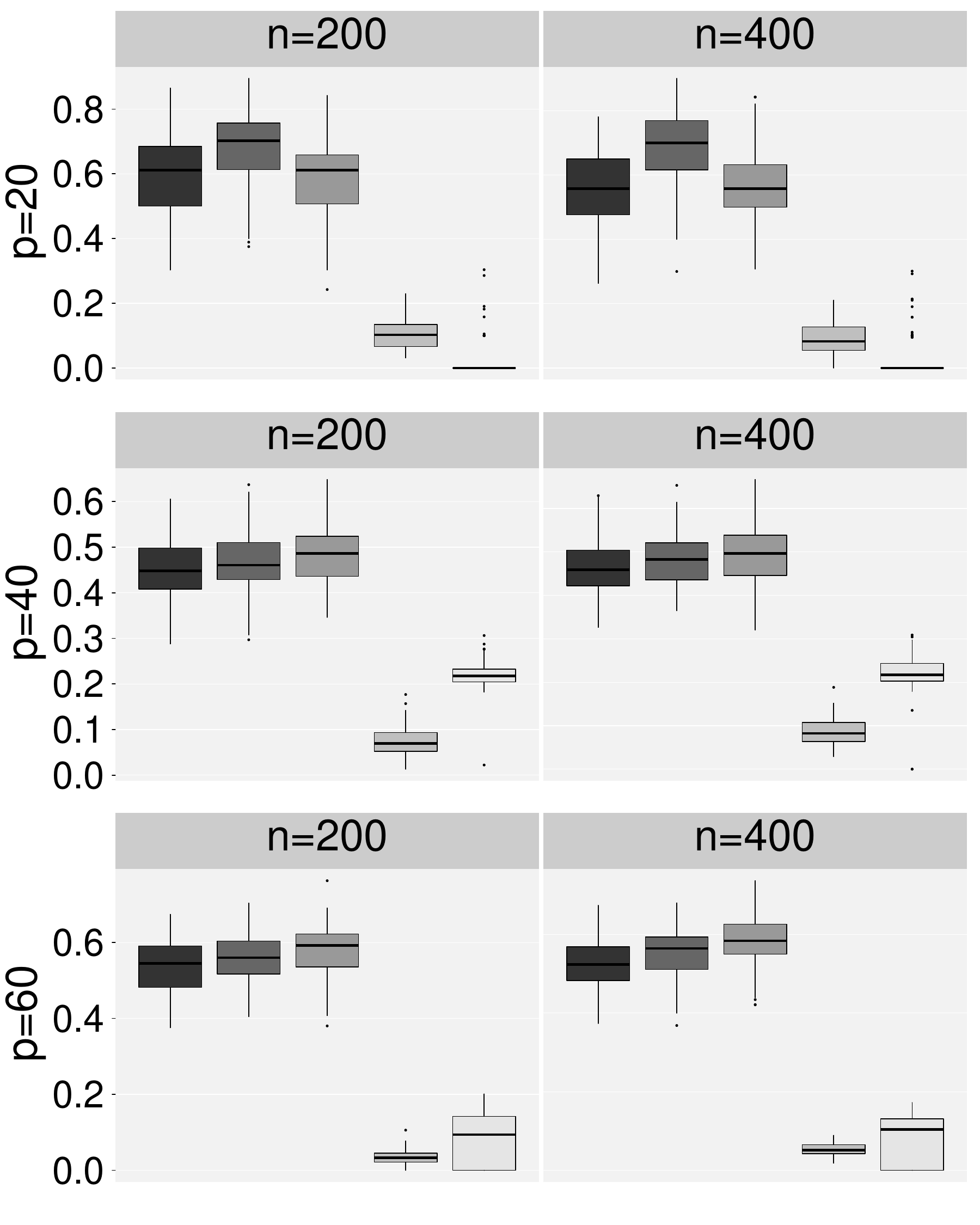}
        \label{fig:sub4_nlngv}
    \end{subfigure}
    \vspace{-1.2cm}
    \caption{Simulation results under model (v); the boxes from left to right represent the five methods, FMixedANM, FLinANM, FNonANM, FNG, and FLNGB.}
    \label{fig:5} 
\end{figure}

By Proposition~\ref{prop_01}, condition \eqref{eq:2.2} holds for the models (i) and (ii), and we show that condition \eqref{eq:2.2} also holds for models (iii) and (iv). In contrast, model (v) violates condition \eqref{eq:2.2}, and thus can be used to assess the robustness of our approach when the proposed identifiability condition fails to hold. Detailed calculations whether condition \eqref{eq:2.2} holds or not for models (iii), (iv) and (v) are provided in Section~E.3 of the Supplementary Material.

To evaluate the performance of each method, we compute two metrics, the structural Hamming distance (SHD) and the F1 score. Letting $\mathbb E^0$ denote the true edge set and $\hat{\mathbb E}$ denote the estimated edge set from any given method, we define
\begin{align*}
\mathrm{SHD} = \mathrm{FP} + \mathrm{FN} - R,~~\mathrm{F1} = \frac{2\,\mathrm{TP}}{2\,\mathrm{TP}+\mathrm{FP}+\mathrm{FN}},
\end{align*}
where $\mathrm{TP} = |\{(i,j) \in \hat{\mathbb E} : (i,j) \in \mathbb E^0\}|$, $\mathrm{FP} = |\{(i,j) \in \hat{\mathbb E} : (i,j) \notin \mathbb E^0\}|$, $\mathrm{FN} = |\{(i,j) \in \mathbb E^0 : (i,j) \notin \hat{\mathbb E}\}|$, and $R = |\{(i,j) \in \hat{\mathbb E} : (j,i) \in \mathbb E^0\}|$ counts the estimated edges whose reversal appears in the true DAG. When $2\,\mathrm{TP}+\mathrm{FP}+\mathrm{FN}$ is zero, we define $\mathrm{F1}=0$ by convention. A smaller SHD and a larger F1 score indicate a better DAG reconstruction.

The simulation results under models (i), (iii) and (v) are presented in Figures~\ref{fig:1} to \ref{fig:5}; Figures corresponding to models (ii) and (iv) are put in Section E.4 of the Supplementary Material for the sake of space. Under the linear models (i) and (ii) in terms of the F1 score, FMixedANM performs the best, closely followed by FLinANM; FNonANM and FLNGB perform comparably, whereas FNG, which is not consistent under linear models, performs the worst. In terms of the SHD, FLNGB performs the best, followed by three proposed methods.

Under the non-linear models (iii) to (v), the advantage of our proposed methods is significant: In terms of both the SHD and the F1 score, all the proposed methods outperform FNG, which, under models (iii) and (iv), is better than FLNGB that is not consistent under non-linear models. Recall that FLinANM is not consistent under non-linear models and condition~\eqref{eq:2.2} is violated in model (v), these results indicate promising robustness of our methodology.
 
Within our proposed methods, FMixedANM almost always ranks among the best two and indeed achieves the best performance under a few settings. These observations show considerable usefulness of our mixed regression model serving as a robust default across linear and non-linear models. 

\subsection{Brain effective connectivity analysis}\label{subsc_real}

We illustrate the proposed methods through a brain effective
connectivity analysis of an electroencephalography (EEG) dataset. The data were collected from a study of genetic predisposition to alcoholism conducted at the Neurodynamics Laboratory of the State University of New York Health Center \citep{Zhang1995} available at \citet{UCIeeg}, and have been analyzed as multivariate functional data by \citet{Zhu2016} and \citet{Qiao2019}. The study consists of $n=122$ subjects, among which $77$ subjects are in the alcoholic group and $45$ subjects are in the control group. During the experiment, each subject was exposed to either a single stimulus or two stimuli presented in a matched or non-matched
condition, and completed $120$ trials in total. The EEG activity was recorded from $p=64$ electrodes placed at standard locations on the subject's scalp, sampled at $256$ Hz for one second after each stimulus, which yielded $r=256$ time points per trial. Similar to \citet{Li2010} and \citet{Qiao2019}, we average the trial-specific signals for each subject and electrode without band-pass filtering, and treat the resulting mean curves as independent functional observations. Trials with corrupted files are excluded. This results in $n=77$ functional observations for the alcoholic group and $n=45$ for the control group, each involving $p=64$ functions observed at $256$ time points. We apply all five methods in Section~\ref{sc_sim} to the two groups separately. 

Table~S1 in Section E.4 of the Supplementary Material reports the regional distribution of the group-specific edges for all methods, i.e., the directed edges identified in only one of the two
groups, with each edge $i\rightarrow j$ counted in the region of its source electrode $i$. We can see overall consistent regional
patterns among different methods in Table S1. Specifically, parietal electrodes contribute the largest or second-largest number of alcohol-specific edges for the three proposed methods---FMixedANM ($12$ of $26$), FLinANM ($15$ of $58$) and FNonANM ($10$ of $24$)---and remain a major source for FLNGB ($27$ of $164$), whereas FNG concentrates these edges at frontal-pole electrodes (FP1, FP2 and FPZ). Frontal electrodes contribute the most control-specific edges for each method:
FMixedANM ($7$ of $14$), FLinANM ($14$ of $44$), FNonANM ($7$ of $7$), FNG ($16$ of $28$) and FLNGB ($22$ of $83$). The largest discrepancy concerns FNG and FLNGB, which place a fraction of their edges in occipital regions and the non-cortical
channels (X, Y and nd).

\begin{figure}[t!]
  \centering
  \vspace{-2cm}
\makebox[\textwidth][c]{%
    \includegraphics[width=1.05\textwidth]{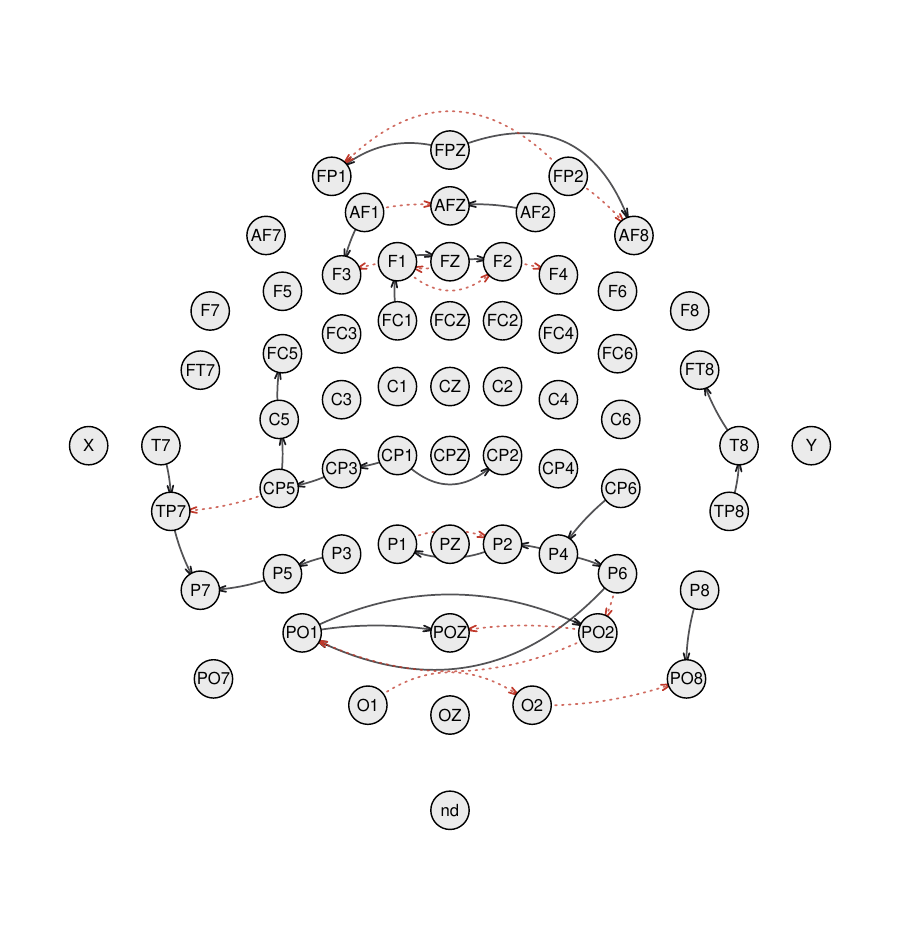}%
  }
  \vspace{-2.5cm}
  \caption{Differential DAGs of the EEG dataset estimated by
  FMixedANM. Solid (dashed) lines denote edges identified only by the alcoholic (control) group. }
  \label{fig:diff}
\end{figure} 

To investigate the difference between the two groups, we next compare the edge sets estimated by FMixedANM. Among the $36$ edges identified in the alcoholic group and the $24$ edges identified in the control group, $10$ edges are shared by both groups, $26$ edges involving
$35$ electrodes are identified only by the alcoholic group, and $14$ edges involving $21$ electrodes are identified only by the control group. Figure~\ref{fig:diff} plots the group-specific differential edges, where solid (dashed) lines denote edges identified only by the alcoholic (control) group.

The alcohol-specific edges are concentrated in the parietal region: $12$ of the $26$ alcohol-specific edges originate from parietal
electrodes. In contrast, the
control-specific edges are predominantly frontal ($7$ of $14$ originate from frontal electrodes) with the
remaining edges located in the parietal and occipital regions. These observations are consistent with the existing literature on this dataset \citep{Zhu2016,Qiao2019} as well as  medical studies of EEG connectivity in alcoholism. Specifically, \citet{Bae2017} found altered effective connectivity estimated from scalp EEG in alcoholism.  \citet{Meyers2020} showed that genetic liability to alcohol dependence is associated with elevated fronto-central, temporo-parietal and parietal-occipital
EEG connectivity. These regional patterns are broadly consistent with previous EEG studies. However, a single small-sample analysis does not validate the inferred causal directions.

\section{Discussion}\label{sc_dis}
In this work, we propose a two-stage procedure for functional causal discovery that first estimates a causal ordering by sequentially minimizing conditional covariance norms and then recovers the directed edges through a SCAD-penalized function-on-function regression. Our identifiability condition does not postulate specific structural or distribution assumptions, and the developed functional regression properly incorporates both linear and non-linear components. We establish the consistency of order determination, regression-coefficient estimation, and edge recovery. Numerical experiments empirically support our methodology.

Two theoretical questions may deserve future investigation. First, the present theory does not account for smoothing or discretization errors, which may be unrealistic under some applications. Second, the number of nodes $p$ is treated as fixed in asymptotic theory. Yet, for the purpose of regression only, the developed FMANM conceptually allows for diverging $p$. Therefore, a rigorous theoretical justification of FMANM under diverging $p$ seems to be interesting.

Beyond theoretical investigations, several methodological and practical extensions deserve further exploration. One direction is to develop joint estimation procedures for multiple related functional DAGs, allowing shared and group-specific causal relationships across populations. Another is to extend the framework to heterogeneous data containing both functional and vector-valued variables, thereby accommodating applications in which trajectories are analyzed together with demographic, clinical, or environmental measurements. From a practical perspective, incorporating scientific knowledge or available interventional data into graph estimation, and validating the inferred relationships using independent datasets or experimental techniques would strengthen the practical relevance and scientific interpretation of functional causal discovery.

\section*{Disclosure Statement}
ChatGPT 6 has been used to check mathematical details and polish text. The authors report there are no competing interests to declare.

\bibliography{ref}  

\end{document}